\documentclass[journal]{IEEEtran}

\usepackage[english]{babel}								
\makeatletter
\adddialect\l@ENGLISH\l@english
\makeatother

\usepackage[T1]{fontenc}

\usepackage{mathtools}
\usepackage{amssymb, amsthm, amscd, MnSymbol}
\DeclareMathAlphabet{\mathbbm}{U}{bbm}{m}{n}
\usepackage{bm}
\usepackage{nicefrac}

\usepackage{algorithm,algorithmic}
\makeatletter
\@addtoreset{algorithm}{section}
\makeatother

\ifCLASSOPTIONcompsoc
  \usepackage[caption=false,font=normalsize,labelfont=sf,textfont=sf]{subfig}
\else
  \usepackage[caption=false,font=footnotesize]{subfig}
\fi

\usepackage{float}
\usepackage{fixltx2e}
\usepackage{stfloats}

\usepackage{cite}

\usepackage[pdftex]{graphicx}	
\usepackage[none]{hyphenat}
\usepackage{wrapfig} 
\usepackage{pgfplots}
\DeclareMathOperator{\Tr}{tr}

\newtheorem{theorem}{Theorem}[section]
\newtheorem{definition}[theorem]{Definition}

\newtheorem{proposition}[theorem]{Proposition}
\newtheorem{corollary}[theorem]{Corollary}

\numberwithin{equation}{section}						
\numberwithin{figure}{section}							
\numberwithin{table}{section}							
\begin{document}

\title{Spectral Super-resolution With Prior Knowledge}

\author{Kumar Vijay~Mishra,
        Myung~Cho,
        Anton~Kruger,       
        and~Weiyu~Xu
\thanks{The authors are with the Department of Electrical and Computer Engineering, The University of Iowa, Iowa City, IA, 52242 USA, e-mail: \{kumarvijay-mishra, myung-cho, anton-kruger, weiyu-xu\}@uiowa.edu}
\thanks{Part of this work has been previously presented in 2014 IEEE International Conference on Acoustics, Speech, and Signal Processing (ICASSP).}}

\maketitle

\begin{abstract}
We address the problem of super-resolution frequency recovery using prior knowledge of the structure of a spectrally sparse, undersampled signal. In many applications of interest, some structure information about the signal spectrum is often known. The prior information might be simply knowing precisely some signal frequencies or the likelihood of a particular frequency component in the signal. We devise a general semidefinite program to recover these frequencies using theories of positive trigonometric polynomials. Our theoretical analysis shows that, given sufficient prior information, perfect signal reconstruction is possible using signal samples no more than thrice the number of signal frequencies. Numerical experiments demonstrate great performance enhancements using our method. We show that the nominal resolution necessary for the grid-free results can be improved if prior information is suitably employed.
\end{abstract}


\IEEEpeerreviewmaketitle

\section{Introduction}
\label{sec:intro}
\IEEEPARstart{I}{n} many areas of engineering, it is desired to infer the spectral contents of a measured signal. In the absence of any \textit{a priori} knowledge of the underlying statistics or structure of the signal, the choice of spectral estimation technique is a subjective craft \cite{marple1987digital, stoica2005spectral}. However, in several applications, the knowledge of signal characteristics is available through previous measurements or prior research. By including such prior knowledge during spectrum estimation process, it is possible to enhance the performance of spectral analysis.

One useful signal attribute is its sparsity in spectral domain. In recent years, spectral estimation methods that harness the spectral sparsity of signals have attracted considerable interest \cite{mishali2010theory, duarte2013spectral, candes2013towards, tang2012csotg}. These methods trace their origins to \textit{compressed sensing} (CS) that allows accurate recovery of signals sampled at sub-Nyquist rate \cite{donoho2006compressed}. In the particular context of spectral estimation, the signal is assumed to be sparse in a finite discrete dictionary such as Discrete Fourier Transform (DFT). As long as the true signal frequency lies in the center of a DFT bin, the discretization in frequency domain faithfully represents the continuous reality of the true measurement. If the true frequency is not located on this discrete frequency grid, then the aforementioned assumption of sparsity in the DFT domain is no longer valid \cite{tan2014sparse, huang2012adaptive}. The result is an approximation error in spectral estimation often referred to as scalloping loss \cite{harris1978use}, basis mismatch \cite{chi2011sensitivity}, and gridding error \cite{fannjiang2012coherence}.

Recent state-of-the-art research \cite{candes2013towards, tang2012csotg, tang2013justdiscretize} has addressed the problem of basis mismatch by proposing compressed sensing in continuous spectral domain. This \textit{grid-free} approach is inspired by the problems of total variation minimization \cite{candes2013towards} and atomic norm minimization \cite{tang2012csotg} to recover \textit{super-resolution} frequencies - lying anywhere in the continuous domain $[0, 1]$ - with few random time samples of the spectrally sparse signal, provided the line spectrum maintains a nominal separation. A number of generalizations of off-the-grid compressed sensing for specific signal scenarios have also been attempted, including extension to higher dimensions \cite{chi2013robust, chi2013compressive, xu2013precise}.

However, these formulations of off-the-grid compressed sensing assume no prior knowledge of signal other than sparsity in spectrum. In fact, in many applications, where signal frequencies lie in continuous domain such as radar \cite{skolnik2008radar}, acoustics \cite{trivett1981modified}, communications \cite{beygi2014multiscale}, and power systems \cite{zygarlicki2012prony}, additional prior information of signal spectrum might be available. For example, a radar engineer might know the characteristic speed with which a fighter aircraft flies. This knowledge then places the engineer in a position to point out the ballpark location of the echo from the aircraft in the Doppler frequency spectrum. Similarly, in a precipitation radar, the spectrum widths of echoes from certain weather phenomena (tornadoes or severe storms) are known from previous observations \cite{doviak1993doppler}. This raises the question whether we can use signal structures beyond sparsity to improve the performance of spectrum estimation.

There are extensive works in compressed sensing literature that discuss recovering sparse signals using secondary signal support structures, such as \textit{structured sparsity} \cite{cevher2009recovery} (tree-sparsity \cite{baraniuk2010model}, block sparsity \cite{stojnic2009reconstruction}, and Ising models \cite{cevher2008sparse}), spike trains \cite{hegde2009compressive, azais2013spike}, nonuniform sparsity \cite{amin2009weighted, vaswani2010modified},  and multiple measurement vectors (MMVs) \cite{duarte2011structured}. However, these approaches assume discrete-valued signal parameters while, in the spectrum estimation problem, frequencies are continuous-valued. Therefore, the techniques of using prior support information in discrete compressed sensing for structured sparsity do not directly extend to spectrum estimation. Moreover, it is rather unclear as to how general signal structure constraints can be imposed for super-resolution recovery of continuous-valued frequency components.

In this paper, we focus on a more generalized approach to super-resolution that addresses the foregoing problems with line spectrum estimation. We propose continuous-valued line spectrum estimation of irregularly undersampled signal in the presence of structured sparsity. Prior information about the signal spectrum comes in various forms. For example, in the spectral information concerning a rotating mechanical system, the frequencies of the supply lines or interfering harmonics might be precisely known \cite{wirfalt2011subspace}. However, in a communication problem, the engineer might only know the frequency band in which a signal frequency is expected to show up. Often the prior knowledge is not even specific to the level of knowing the frequency subbands precisely. The availability of previous measurements, such as in remote sensing or bio-medicine, can aid in knowing the likelihood of having an active signal frequency in the neighborhood of a specific spectral band. In this paper, we greatly broaden the scope of prior information that can range from knowing only the likelihood of occurrence of frequency components in a spectral subband to exactly knowing the location of some of the frequencies. 

In all these cases, we propose a precise semidefinite program to perfectly recover all the frequency components. When some frequencies are \emph{precisely} known, we propose to use \textit{conditional atomic norm} minimization to recover the off-the-grid frequencies. In practice, the frequencies are seldom \textit{precisely} known. However, as long as the frequency locations are approximately known to the user, we show that the spectrally sparse signal could still be perfectly reconstructed. Here, we introduce \textit{constrained atomic norm} minimization that accepts the \textit{block priors} - frequency subbands in which true spectral contents of the signal are known to exist - in its semidefinite formulation. When only the probability density function of signal frequencies is known, we incorporate such a \textit{probabilistic prior} in the spectral estimation problem by suggesting the minimization of \textit{weighted atomic norm}. The key is to transform the dual of atomic norm minimization to a semidefinite program using linear matrix inequalities (LMI). These linear matrix inequalities are, in turn, provided by theories of positive trigonometric polynomials \cite{fejer1915uber}. Our methods boost the signal recovery by admitting lesser number of samples for spectral estimation and decreasing reliance on the minimum resolution necessary for super-resolution. If the prior information locates the frequencies within very close boundaries of their true values, then we show that it is possible to perfectly recover the signal using samples no more than thrice the number of signal frequencies.

Our work has close connections with a rich heritage of research in spectral estimation. For uniformly sampled or \textit{regularly spaced} signals, there are a number of existing approaches for spectral estimation by including known signal characteristics in the estimation process. The classical Prony's method can be easily modified to account for known frequencies \cite{trivett1981modified}. Variants of the subspace-based frequency estimation methods such as MUSIC (MUltiple SIgnal Classification) and ESPRIT (Estimation of Signal Parameters via Rotation Invariance Techniques) have also been formulated \cite{linebarger1995incorporating, wirfalt2011subspace}, where prior knowledge can be incorporated for parameter estimation. For applications wherein only approximate knowledge of the frequencies is available, the spectral estimation described in \cite{zachariah2013line} applies circular von Mises probability distribution on the spectrum. 

For \textit{irregularly spaced} or non-uniformly sampled signal, sparse signal recovery methods which leverage on prior information have recently gained attention \cite{amin2009weighted, vaswani2010modified, ji2008bayesian, bourguignon2007sparsity}. Compressed sensing with clustered priors was addressed in \cite{yu2012bayesian} where the prior information on the number of clusters and the size of each cluster was assumed to be unknown. In \cite{fannjiang2011music}, MUSIC was extended to undersampled, irregularly spaced sparse signals in a discrete dictionary, while \cite{liao2014music} analyzed the performance of \textit{snapshot}-MUSIC for uniformly sampled signals in a continuous dictionary. Our technique is more general; it applies to irregularly sampled signals in a continuous dictionary, and is, therefore, different from known works on utilizing prior information for spectral estimation of regularly sampled signals.

\section{Problem Formulation}
\label{sec:prob_formulation}
In general, the prior information can be available for any of the signal parameters such as amplitude, phase or frequencies. However, in this paper, we restrict the available knowledge to only the frequencies of the signal. We assume that the amplitude and phase information of any of the spectral component is not known, irrespective of the pattern of known frequency information. Our approach is to first analyze the case of a more nebulous prior information, that is the probabilistic priors, followed by an interesting special case of block priors. The case when \textit{some} frequencies are precisely known is considered in the end where, unlike previously considered cases, we recover the signal using the semidefinite program for the primal problem.

We consider a frequency-sparse signal $x[l]$ expressed as a sum of $s$ complex exponentials,\small
\begin{equation}
\label{eq:sigmodelstd}
x[l] = \sum\limits_{j=1}^{s} c_je^{i2\pi f_jl} = \sum\limits_{j=1}^{s} |c_j|a(f_j, \phi_j)[l]\phantom{1}, \phantom{1} l \in \mathcal{N},
\end{equation}\normalsize
where $c_j = |c_j|e^{i\phi_j}$ ($i = \sqrt{-1}$) represents the complex coefficient of the frequency $f_j \in [0, 1]$, with amplitude $|c_j| > 0$, phase $\phi_j \in [0, 2\pi)$, and frequency-\textit{atom} $a(f_j, \phi_j)[l] = e^{i(2\pi f_j l + \phi_j)}$. We use the index set $\mathcal{N} = \{l\phantom{1}|\phantom{1} 0 \le l \le n-1\}$, where $|\mathcal{N}| = n, n \in \mathbb{N}$, to represent the time samples of the signal. We further suppose that the signal in (\ref{eq:sigmodelstd}) is observed on the index set $\mathcal{M} \subseteq \mathcal{N}$, $|\mathcal{M}| = m \leq n$ where $m$ observations are chosen uniformly at random. Our objective is to recover all the continuous-valued the frequencies with very high accuracy using this undersampled signal.

The signal in (\ref{eq:sigmodelstd}) can be modeled as a positive linear combination of the unit-norm frequency-\textit{atoms} $a(f_j, \phi_j)[l] \in \mathcal{A} \subset \mathbb{C}^n$ where $\mathcal{A}$ is the set of all the frequency-atoms. These frequency atoms are basic units for synthesizing the frequency-sparse signal. This leads to the following formulation of the \textit{atomic norm} $||\hat{x}||_\mathcal{A}$ - a sparsity-enforcing analog of $\ell_1$ norm for a general atomic set $\mathcal{A}$:\small
\begin{equation}
\label{eq:atomicnorm}
||\hat{x}||_{\mathcal{A}} = \underset{c_j, f_j}{\text{inf}}\phantom{1}\left\{\sum\limits_{j=1}^s|c_j|: \hat{x}[l] = \sum\limits_{j=1}^{s} c_je^{i2\pi f_jl} \phantom{1}, \phantom{1} l \in \mathcal{M}\right\}.
\end{equation}\normalsize

To estimate the remaining $\mathcal{N} \setminus \mathcal{M}$ samples of the signal $x$, \cite{chandrasekaran2012theconvex} suggests minimizing the atomic norm $||\hat{x}||_\mathcal{A}$ among all vectors $\hat{x}$ leading to the same observed samples as $x$. Intuitively, the atomic norm minimization is similar to $\ell_1$-minimization being the tightest convex relaxation of the combinatorial $\ell_0$-minimization problem. The \textit{primal} convex optimization problem for atomic norm minimization can be formulated as follows,\small
\begin{flalign}
	\label{eq:atomicminimization}
	& \underset{\hat{x}}{\text{minimize}}\phantom{1}  \|\hat{x}\|_{\mathcal{A}}\nonumber\\
	& \text{subject to}\phantom{1} \hat{x}[l] = x[l], \phantom{1} l \in \mathcal{M}.
\end{flalign}\normalsize
Equivalently, the off-the-grid compressed sensing \cite{tang2012csotg} suggests the following semidefinite characterization for $||\hat{x}||_\mathcal{A}$:
\begin{definition} \cite{tang2012csotg} Let $T_n$ denote the $n \times n$ positive semidefinite Toeplitz matrix, $t \in \mathbb{R}^+$, Tr($\cdot$) denote the trace operator and $(\cdot)^*$  denote the complex conjugate. Then,\small
\begin{equation}
\label{eq:atomicnorm_sdpdef}
	||\hat{x}||_{\mathcal{A}} = \underset{T_n, t}{\text{inf}} \left\{\dfrac{1}{2|\mathcal{N}|} \text{Tr($T_n$)} + \frac{1}{2}t : \begin{bmatrix*}[r] T_n & \hat{x} \\ \hat{x}^* & t \end{bmatrix*} \succeq 0 \right\}.
\end{equation}\normalsize
The positive semidefinite Toeplitz matrix $T_n$ is related to the frequency atoms through the following Vandermonde decomposition result by Carath{\`e}odory \cite{cara1911uber}:\small
\begin{align}
\label{eq:vandermonde_decomp_def}
T_n &= URU^*,
\end{align}\normalsize
where \small
\begin{align}
U_{lj} &= a(f_{j}, \phi_{j})[l],\label{eq:freq_vandermonde}\\
                    R &= \text{diag}([b_1, \cdots, b_{r}]).
\end{align}\normalsize
The diagonal elements of $R$ are real and positive, and $r = \text{rank}(T_n)$.
\label{def:stdatomicnorm}
\end{definition}
Consistent with this definition, the atomic norm minimization problem for the frequency-sparse signal recovery can now be formulated as a semidefinite program (SDP) with $m$ affine equality constraints:\small
\begin{flalign}
	\label{eq:semiotg}
	& \underset{T_n, \hat{x}, t}{\text{minimize}}\phantom{1} \dfrac{1}{2|\mathcal{N}|} \text{Tr($T_n$)} + \frac{1}{2}t\nonumber\\
	& \text{subject to}\phantom{1} \begin{bmatrix*}[r] T_n & \hat{x} \\ \hat{x}^* & t \end{bmatrix*} \succeq 0\\
	& \hat{x}[l] = x[l], \phantom{1} l \in \mathcal{M}.\nonumber
\end{flalign}\normalsize
When some information about the signal frequencies is known \textit{a priori}, then our goal is to find a signal vector $\hat{x}$ in (\ref{eq:semiotg}) whose frequencies satisfy additional constraints imposed by prior information. In other words, if $\mathcal{C}$ denotes the set of constraints arising due to prior knowledge of frequencies, then our goal is to find the infimum in (\ref{eq:atomicnorm}) over $f_j \in \mathcal{C}$. 

While framing the problem to harness the prior information, a common approach in compressed sensing algorithms is to replace the classical minimization program with its weighted counterpart \cite{amin2009weighted, vaswani2010modified}. However, signals with continuous-valued frequencies do not lead to a direct application of the weighted $\ell_1$ approach. Rather, such an application leads to a fundamental conundrum: the Vandermonde decomposition of positive semidefinite Toeplitz matrices works for general frequencies wherein the frequency atom in (\ref{eq:freq_vandermonde}) can freely take any frequency and phase values, and it is not clear how to further tighten the positive semidefinite Toeplitz structure to incorporate the known prior information. Thus, it is non-trivial to formulate a computable convex program that can incorporate general prior information to improve signal recovery.

\section{Probabilistic Priors}
\label{sec:prob_priors}
\begin{figure}[!t]
\centering
\includegraphics[width=3.5in]{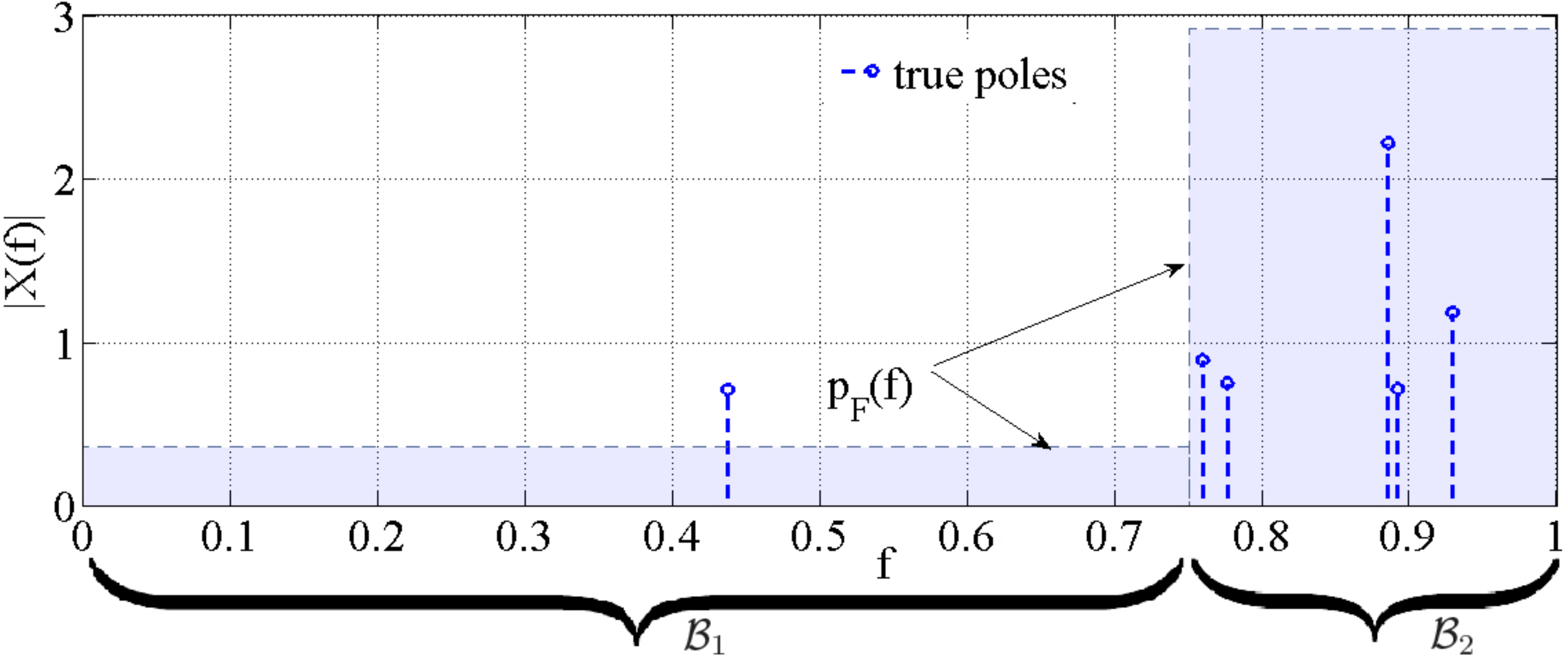}
\caption{\small The probability density function $p_F(f)$ of the frequencies shown with the location of true frequencies in the spectrum $X(f)$ of the signal $x[l]$.}
\label{fig:priorspec}
\end{figure}

In the probabilistic prior model, the probability density function of the frequencies is known. Let $F$ be the random variable that describes the signal frequencies. Let the probability density function (pdf) of F be $p_F(f)$. The problem of line spectrum estimation deals with a finite number of signal frequencies in the domain [0, 1]. For example, we can assume $p_F(f)$ to be piecewise constant as follows. Let the domain $[0,1]$ consist of $p$ disjoint subbands such that $[0, 1] = \bigcup_{k = 1}^{p} \mathcal{B}_k$ where $\mathcal{B}_k$ denotes a subband or a subset of $[0, 1]$. Then the restriction $p_F(f)|_{\mathcal{B}_k}$ of $p_F(f)$ to $\mathcal{B}_k$ is a constant. Figure \ref{fig:priorspec} illustrates a simple case for $p=2$, where the line spectrum $X(f)$ of a signal $x$ is non-uniformly sparse over two frequency subbands $\mathcal{B}_1$ and $\mathcal{B}_2 = [0, 1] \backslash \mathcal{B}_1$, such that the frequencies $f_j$, $j = 1, \cdots, s$, occur in the subinterval $\mathcal{B}_2$ more likely than in $\mathcal{B}_1$.

Intuitively, given probabilistic priors, one may think of recovering the signal $x$ by minimizing a \textit{weighted atomic norm} given by:\small
\begin{equation}
\label{eq:weightedatomicnorm}
||\hat{x}||_{\mathbf{w}\mathcal{A}} = \underset{c_j, f_j}{\text{inf}}\phantom{1}\left\{\sum\limits_{j=1}^sw_j|c_j|: \hat{x}[l] = \sum\limits_{j=1}^{s} c_je^{i2\pi f_j l} \phantom{1}, \phantom{1} l \in \mathcal{M}\right\},
\end{equation}\normalsize
where $\mathbf{w} = \{w_1, \cdots, w_s\}$ is the weight vector, each element $w_j$ of which is associated with the probability of occurrence of the corresponding signal frequency $f_j$. The weight vectors are assigned using a \textit{weight function} $w(f)$. $w(f)$ is a piecewise constant function in the domain $[0, 1]$ such that the restriction $w(f)|_{\mathcal{B}_k}$ of $w(f)$ to $\mathcal{B}_k$ is a constant. Therefore, $\forall \phantom{1} \{f_1, \cdots, f_j\} \in \mathcal{B}_k$, we have $w_1 = \cdots = w_j = w(f)|_{f\in \mathcal{B}_k} = w(f_{\mathcal{B}_k})$ (say). The $w(f)$ is a decreasing function of the sparsity associated with the corresponding frequency subband so that the subband with higher (lower) value of pdf or lesser (more) sparsity is weighted lightly (heavily).

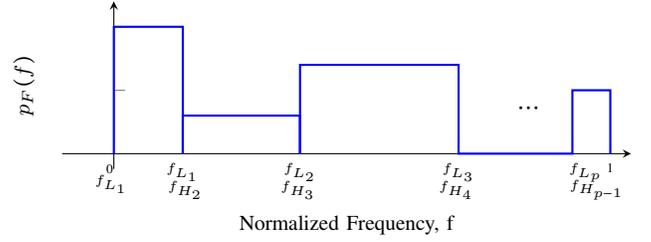
\begin{figure}[!t]
\centering
	\begin{tikzpicture}
	\begin{axis}[
		width=3.6in,
		height=1.5in,
		x axis line style={-stealth},
		y axis line style={-stealth},
		xtick={1.0, 2.7, 5.0, 6.65, 7.2},
		xticklabels={\tiny$f_{L_1}$, \tiny$f_{L_2}$, \tiny$f_{L_3}$, \tiny$\phantom{1}\phantom{1}\phantom{1}f_{L_p}$, \tiny1},
		yticklabels={},
		ymax = 1.2,xmax=7.5,
		axis lines*=center,
		ylabel={\footnotesize $p_F(f)$},
		xlabel={\footnotesize Normalized Frequency, f},
		xlabel near ticks,
		ylabel near ticks]
	\addplot+[thick,mark=none,const plot]
	coordinates
	{(0,0) (0,1) (1.0,1.0) (1.0,0) (1.0,0.3) (2.7,0.3) (2.7,0) (2.7,0.7) (5,0.7) (5,0) (6.65,0) (6.65,0.5) (7.2,0.5) (7.2,0)};
	\end{axis}	
	\node [left] at (0.8,  0) {\tiny0};
    \node [left] at (1.0,  -0.2) {\tiny $f_{L_1}$};
    \node [left] at (2.0,  -0.28) {\tiny $f_{H_2}$};
    \node [left] at (3.5,  -0.28) {\tiny $f_{H_3}$};
    \node [left] at (5.6,  -0.28) {\tiny $f_{H_4}$};
    \node [left] at (7.6,  -0.28) {\tiny $f_{H_{p-1}}$};
	\node [above] at (6.2,  0.6) {$\cdots$};
	\end{tikzpicture}
	\caption{\small The individual frequencies of spectrally parsimonious signal are assumed to lie in known frequency subbands within the normalized frequency domain $[0, 1]$. We assume that all subbands are non-overlapping so that when $f_{H_{k-1}} = f_{L_{k}}$, then $\mathcal{B}_{k-1} = [f_{L_{k-1}}, f_{H_{k-1}}]$ and $\mathcal{B}_{k} = (f_{L_k}, f_{H_k}]$.}
	\label{fig:mult_probbands}
\end{figure}

The problem of line spectral estimation using probabilistic prior can now be presented as the (primal) optimization problem concerning the weighted atomic norm:\small
\begin{flalign}
	\label{eq:primalweighted}
	& \underset{\hat{x}}{\text{minimize}}\phantom{1}  \|\hat{x}\|_{\mathbf{w}\mathcal{A}}\nonumber\\
	& \text{subject to}\phantom{1} \hat{x}[j] = x[j], \phantom{1} l \in \mathcal{M}.
\end{flalign}\normalsize
But we now observe that, unlike weighted $\ell_1$ norm \cite{amin2009weighted}, a semidefinite characterization of the weighted atomic norm does not evidently result from (\ref{eq:semiotg}). Instead, we propose a new semidefinite program for the weighted atomic norm using theories of positive trigonometric polynomials, by looking at its dual problem. For the standard atomic norm minimization problem (\ref{eq:atomicminimization}), the dual problem is framed in this manner:\small
\begin{flalign}
	\label{eq:dualtoatomicminimization}
	 \underset{q}{\text{maximize}} & \phantom{1} \langle q_{\mathcal{M}},x_{\mathcal{M}}\rangle_{\mathbb{R}}\phantom{1}  \nonumber\\
	 \text{subject to}&\phantom{1}\|q\|_{\mathcal{A}}^* \leq 1 \\
	&\phantom{1} q_{\mathcal{N}\setminus \mathcal{M}}=0, \nonumber
\end{flalign}\normalsize
where $\|\cdot\|^*$ represents the dual norm. This dual norm is defined as\small
\begin{align}
\label{eq:dualatomicnorm}
\|q\|_{\mathcal{A}}^*=\sup_{\|\hat{x}\|_{\mathcal{A}}\leq 1} \langle q,\hat{x} \rangle_{\mathbb{R}}=\sup_{f \in [0,1]}|\langle q, a(f,0)\rangle|.
\end{align}\normalsize
For the weighted atomic norm minimization, the primal problem (\ref{eq:primalweighted}) has only equality constraints. As a result, Slater's condition is satisfied and, therefore, strong duality holds \cite{boyd2004convex}. In other words, solving the dual problem also yields an exact solution to the primal problem. The dual of weighted atomic norm is given by\small
\begin{align}
\label{eq:dualweightedatomicnorm}
\|q\|_{\mathbf{w}\mathcal{A}}^*&=\sup_{\|\hat{x}\|_{\mathbf{w}\mathcal{A}}\leq 1} \langle q,\hat{x} \rangle_{\mathbb{R}}=\sup_{\phi \in [0, 2\pi], f \in [0,1]}\Big\langle q, \frac{1}{w(f)}e^{i\phi}a(f,0)\Big\rangle_{\mathbb{R}}\nonumber\\
&=\sup_{f \in [0,1]}\Big|\Big\langle q, \frac{1}{w(f)}a(f,0)\Big\rangle\Big|.
\end{align}\normalsize
The dual problem to (\ref{eq:primalweighted}) can be stated hence,\small
\begin{flalign}
	\label{eq:weighteddual1}
	 \underset{q}{\text{maximize}} & \phantom{1} \langle q_{\mathcal{M}},x_{\mathcal{M}}\rangle_{\mathbb{R}}\phantom{1}  \nonumber\\
	 \text{subject to}&\phantom{1}\|q\|_{\mathbf{w}\mathcal{A}}^* \leq 1 \\
	&\phantom{1} q_{\mathcal{N}\setminus \mathcal{M}}=0, \nonumber
\end{flalign}\normalsize
which by substitution of (\ref{eq:dualweightedatomicnorm}) becomes,\small
\begin{flalign}
	\label{eq:weighteddual2}
	 \underset{q}{\text{maximize}} & \phantom{1} \langle q_{\mathcal{M}},x_{\mathcal{M}}\rangle_{\mathbb{R}}\phantom{1}  \nonumber\\
	 \text{subject to}&\phantom{1}\sup_{f \in [0,1]}\Big|\Big\langle q, \frac{1}{w(f)}a(f,0)\Big\rangle\Big| \leq 1 \\
	&\phantom{1} q_{\mathcal{N}\setminus \mathcal{M}}=0. \nonumber
\end{flalign}\normalsize
Let the probabilistic priors consist of distinct weights for $p$ different frequency subbands $\mathcal{B}_k \subset [0, 1]$, $k = 1, \cdots, p$ such that $[0, 1] = \bigcup_{k = 1}^{p} \mathcal{B}_k = \bigcup_{k = 1}^{p} [f_{L_k}, f_{H_k}]$, where $f_{L_k}$ and $f_{H_k}$ are, respectively, the lower and upper cut-off frequencies for each of the band $\mathcal{B}_k$ (Figure \ref{fig:mult_probbands}). If the probability density function is constant within a frequency band, then the results of the supremums in (\ref{eq:weighteddual2}) need not depend on the weight functions, and therefore, the inequality constraint in the dual problem in (\ref{eq:weighteddual2}) can be expanded as,\small
\begin{flalign}
	\label{eq:weighteddual4}
	 \underset{q}{\text{maximize}} & \phantom{1} \langle q_{\mathcal{M}},x_{\mathcal{M}}\rangle_{\mathbb{R}}\phantom{1}  \nonumber\\
	 \text{subject to}&\phantom{1}\sup_{f \in \mathcal{B}_1}|\langle q, a(f,0)\rangle| \leq w(f_{\mathcal{B}_1}) \nonumber\\
     &\phantom{1}\sup_{f \in \mathcal{B}_2}|\langle q, a(f,0)\rangle| \leq w(f_{\mathcal{B}_2}) \nonumber\\
     &\vdots\nonumber\\
     &\phantom{1}\sup_{f \in \mathcal{B}_p}|\langle q, a(f,0)\rangle| \leq w(f_{\mathcal{B}_p}) \nonumber\\
	&\phantom{1} q_{\mathcal{N}\setminus \mathcal{M}}=0.
\end{flalign}\normalsize
We now map each of the inequality constraints in the foregoing dual problem to a linear matrix inequality, leading to the semidefinite characterization of the weighted atomic norm minimization. We recognize that the constraints in (\ref{eq:weighteddual4}) imply $Q(f) = \langle q, a(f,0)\rangle$ is a positive trigonometric polynomial \cite{fejer1915uber} in $f \in \mathcal{B}_k$, since\small
\begin{align}
\label{eq:dualpoly_expression}
Q(f) = \langle q, a(f,0) \rangle = \sum\limits_{l=0}^{n-1} q_{l}e^{-i 2\pi f l }.
\end{align}\normalsize
Such a polynomial can be parameterized by a particular type of positive semidefinite matrix. Thus, we can transform the polynomial inequality, such as the ones in (\ref{eq:weighteddual4}), to a linear matrix inequality.

\subsection{Gram Matrix Parametrization}
\label{subsec:gram_ptp}
A trigonometric polynomial $R(z) = \sum\limits_{k=-(n-1)}^{n-1}r_k z^{-k}$, which is also nonnegative on the entire unit circle, can be parametrized using a positive semidefinite, Hermitian matrix $\bm{G}$ (called the \textit{Gram} matrix) that identifies the polynomial coefficients $r_k$ as a function of its elements \cite[p. 23]{dumitrescu2007positive}:
\begin{align}
\label{eq:gram_parmet}
r_k &= tr[\mathbf{\Theta}_k \bm{G}],
\end{align}
where $\bm{\Theta}_k$ is an elementary Toeplitz matrix with ones on its $k$th diagonal and zeros elsewhere. Here, $k = 0$ corresponds to the main diagonal, and $k$ takes positive and negative values for upper and lower diagonals respectively.

For the trigonometric polynomial that is nonnegative only over an arc of the unit circle, we have the following theorem:
\begin{theorem} \cite[p. 12]{dumitrescu2007positive} A trigonometric polynomial\small
\begin{align}
\label{eq:expn_ptp}
R(z) &= \sum\limits_{k=-(n-1)}^{n-1}r_k z^{-k},\phantom{1} r_{-k} = r_k^*,
\end{align}\normalsize
where $R \in \mathbb{C}_{n-1}[z]$ for which $R(\omega) \ge 0$, for any $z = e^{i\omega}$, $\omega \in [\omega_L, \omega_H] \subset [-\pi, \pi]$, can be expressed as\small
\begin{align}
R(z) = F(z)F^*(z^{-1}) + D_{\omega_L\omega_H}(z).G(z)G^*(z^{-1}),
\end{align}\normalsize
where $F(z)$, and $G(z)$ are causal polynomials with complex coefficients, of degree at most $n-1$ and $n-2$, respectively. The polynomial\small
\begin{align}
D_{\omega_L\omega_H}(z) &= d_1z^{-1} + d_0 + d_1^*z
\end{align}\normalsize
where\small
\begin{align}
d_0 &= -\dfrac{\alpha\beta+1}{2}\\
d_1 &= \dfrac{1-\alpha\beta}{4} + j\dfrac{\alpha+\beta}{4}\label{eq:d1_exp}\\
\alpha &= \tan{\dfrac{\omega_L}{2}}\\
\beta &= \tan{\dfrac{\omega_H}{2}},
\end{align}\normalsize
is defined such that $D_{\omega_L\omega_H}(\omega)$ is nonnegative for $\omega \in [\omega_L, \omega_H]$ and negative on its complementary.\footnote{cf. Errata to \cite{dumitrescu2007positive} available online. The 2007 print edition of \cite{dumitrescu2007positive} has an error in the expression (\ref{eq:d1_exp}).}
\label{thm:dumitrescu_1_15}
\end{theorem}
Since $F(z)$ and $G(z)$ are causal polynomials, the products $F(z)F^*(z^{-1})$ and $G(z)G^*(z^{-1})$ are positive trigonometric polynomials that can each be separately parameterized with Gram matrices $\bm{G}_1$ and $\bm{G}_2$ respectively.
\begin{proposition}
A trigonometric polynomial $R$ in (\ref{eq:expn_ptp}) that is nonnegative on the arc $[\omega_L, \omega_H] \subset [-\pi, \pi]$ or, alternatively, the subband $[f_L, f_H] \subset [0, 1]$, can be parameterized using the Gram matrices  $\bm{G}_1 \in \mathbb{C}^{n \times n}$ and $\bm{G}_2 \in \mathbb{C}^{(n-1) \times (n-1)}$ as follows:\small
\begin{align}
\label{eq:lmi_freq}
r_k &= tr[\mathbf{\Theta}_k \bm{G}_1] + \Tr{[(d_1\mathbf{\Theta}_{k-1} + d_0\mathbf{\Theta}_{k} + d_1^*\mathbf{\Theta}_{k+1}) \cdot \bm{G}_2]}\nonumber\\
	&\triangleq \mathcal{L}_{k, f_L, f_H}(\bm{G}_1, \bm{G}_2),
\end{align}\normalsize
where we additionally require the elementary Toeplitz matrix $\mathbf{\Theta}_k$ in the second argument to be a nilpotent matrix of order $n-k$ for $|k|>0$. The translation of frequencies between the two domains is given by:\small
\begin{align}
\omega_L &= \begin{dcases*}
        			2\pi f_L & : 0 $\le$ $f_L$ $\le$ 0.5\\
        			2\pi(f_L - 1) & : 0.5 < $f_L$ < 1
        	   \end{dcases*}\\
\omega_H &= \begin{dcases*}
        			2\pi f_H & : 0 < $f_H$ $\le$ 0.5\\
        			2\pi(f_H - 1) & : 0.5 < $f_H$ $\le$ 1
        	   \end{dcases*}.
\end{align}\normalsize
\label{prop:lmi_bandedpoly}
\end{proposition}
\begin{IEEEproof}[Proof of Proposition \ref{prop:lmi_bandedpoly}.] Let $F(z)$ and $G(z)$ be causal polynomials such that, $F(z)$ $=$ $\mathbf{f}^T\psi(z^{-1})$, and $G(z)$ $=$ $\mathbf{g}^T\phi(z^{-1})$, where $\mathbf{f}$ $=$ $\begin{bmatrix}f_0 & f_1 & \cdots & f_{n-1}\end{bmatrix}^T$ $\in$ $\mathbb{C}^n$, and $\mathbf{g}$ $=$ $\begin{bmatrix}g_0 & g_1 & \cdots & g_{n-2}\end{bmatrix}^T$ $\in$ $\mathbb{C}^{n-1}$ are vectors of coefficients of the causal polynomials $F(z)$ and $G(z)$ respectively, and $\psi(z^{-1})$ $=$ $\begin{bmatrix}1 & z^{-1} & ... & z^{-(n-1)} \end{bmatrix}^T$, and $\phi(z^{-1})$ $=$ $\begin{bmatrix}1 & z^{-1} & ... & z^{-(n-2)} \end{bmatrix}^T$, are the canonical basis vectors of the corresponding polynomials. Let\small
\begin{align}
R(z) &= \sum\limits_{k=-(n-1)}^{(n-1)}r_k z^{-k},\phantom{1} r_{-k} = r_k^*\nonumber\\
A(z) &= \sum\limits_{k=-(n-1)}^{n-1}a_k z^{-k} = F(z)F^*(z^{-1}),\phantom{1} a_{-k} = a_k^*\nonumber\\
B(z) &= \sum\limits_{k=-(n-2)}^{n-2}b_k z^{-k} = G(z)G^*(z^{-1}),\phantom{1} b_{-k} = b_k^*\nonumber\\
\tilde{B}(z) &= \sum\limits_{k=-(n-1)}^{n-1}\tilde{b}_k z^{-k} = D_{\omega_L\omega_H}(z).G(z)G^*(z^{-1}),\phantom{1} \tilde{b}_{-k} = \tilde{b}_k^{*}.\nonumber
\end{align}\normalsize
From the above, $r_k = a_k + \tilde{b}_k$. Let $\bm{G}_1 \in \mathbb{C}^{n \times n}$ and $\bm{G}_2 \in \mathbb{C}^{(n-1) \times (n-1)}$ be the Gram matrices. Then, as shown in (\ref{eq:gram_parmet}), the parameterization process yields, $a_k = tr[\mathbf{\Theta}_k \bm{G}_1]$. Also, by definition, if the Gram matrix $\bm{G}_2$ is associated with a trigonometric polynomial $B(z)$, then we have\small
\begin{align}
B(z) &= \phi^T(z^{-1}) \cdot \bm{G}_2 \cdot \phi(z) = \Tr{[\phi(z) \cdot \phi^T(z^{-1}) \cdot \bm{G}_2]}\nonumber\\
&= \Tr{[\Phi(z) \cdot \bm{G}_2]},\label{eq:poly_paramet1}
\end{align}\normalsize
where
\begin{align}
\Phi(z) = \begin{bsmallmatrix} 1\\ z\\ \vdots \\ z^{n-2}\end{bsmallmatrix} \begin{bsmallmatrix}1 & z^{-1} & ... & z^{-(n-2)} \end{bsmallmatrix} = \begin{bsmallmatrix} 1 & z^{-1} & \cdots & z^{-(n-2)}\\ z & 1 & \ddots & z^{-(n-3)}\\ \vdots & \ddots & \ddots & \vdots\\ z^{n-2} & z^{n-3} & \cdots & 1 \end{bsmallmatrix}.\nonumber
\end{align}
This leads to the following expressions:\small
\begin{align}
\Phi(z) &= \sum\limits_{k = -(n-2)}^{n-2} \mathbf{\Theta}_k z^{-k},\label{eq:poly_paramet2}\\
z^{-1}\Phi(z) &= z^{-1}\sum\limits_{k = -(n-2)}^{n-2} \mathbf{\Theta}_k z^{-k} = \sum\limits_{k = -(n-3)}^{n-1} \mathbf{\Theta}_{k-1} z^{-k},\label{eq:poly_paramet3}\\
z\Phi(z) &= z\sum\limits_{k = -(n-2)}^{n-2} \mathbf{\Theta}_k z^{-k} = \sum\limits_{k = -(n-1)}^{n-3} \mathbf{\Theta}_{k+1} z^{-k}.\label{eq:poly_paramet4}
\end{align}\normalsize
Substitution of (\ref{eq:poly_paramet2})-(\ref{eq:poly_paramet4}) in (\ref{eq:poly_paramet1}) gives the following matrix-parametric expression,\small
\begin{align}
\begin{aligned}
&\tilde{B}(z) = (d_1z^{-1} + d_0 + d_1^*z)\Tr{[\Phi(z) \cdot \bm{G}_2]}\nonumber\\
&= \Tr{[(d_1z^{-1}\Phi(z) + d_0\Phi(z) + d_1^*z\Phi(z)) \cdot \bm{G}_2]}\nonumber\\
&= \Tr{}[(d_1\sum\limits_{k = -(n-3)}^{n-1} \mathbf{\Theta}_{k-1} z^{-k} + d_0\sum\limits_{k = -(n-2)}^{n-2} \mathbf{\Theta}_{k} z^{-k}\nonumber\\
& + d_1^*\sum\limits_{k = -(n-1)}^{n-3} \mathbf{\Theta}_{k+1} z^{-k}) \cdot \bm{G}_2]\nonumber\\
&= \sum\limits_{k = -(n-1)}^{n-1}\Tr{[(d_1\mathbf{\Theta}_{k-1} + d_0\mathbf{\Theta}_{k} + d_1^*\mathbf{\Theta}_{k+1}) \cdot \bm{G}_2]}z^{-k}.\nonumber
\end{aligned}
\end{align}\normalsize
Then,
\begin{align}
\tilde{b}_k &= \Tr{[(d_1\mathbf{\Theta}_{k-1} + d_0\mathbf{\Theta}_{k} + d_1^*\mathbf{\Theta}_{k+1}) \cdot \bm{G}_2]}.\label{eq:poly_parametB}
\end{align}
Substitution of matrix parameterizations of $a_k$ and $\tilde{b}_k$ in the expression of $r_k$ completes the proof.
\end{IEEEproof}

The dual polynomial $Q(f)$ in (\ref{eq:dualpoly_expression}) is nonnegative on multiple non-overlapping intervals, and can therefore be parameterized by as many different pairs of Gram matrices $\{\bm{G}_1$, $\bm{G}_2\}$ as the number of subbands $p$. In the following subsection, we relate this parametrization to the corresponding probabilistic weights of the subbands.

\subsection{SDP Formulation}
\label{subsec:SDPFormulation}
Based on the Bounded Real Lemma \cite[p. 127]{dumitrescu2007positive} (which, in turn, is based on Theorem \ref{thm:dumitrescu_1_15}), a positive trigonometric polynomial constraint of the type $|R(\omega)| \leq 1$ can be expressed as a linear matrix inequality \cite[p. 143]{dumitrescu2007positive}. Stating this result for the dual polynomial constraint over a single frequency band, such as those in (\ref{eq:weighteddual4}), we have \small
\begin{align}
\label{eq:dualpolyconstraint1}
\sup_{f \in [f_L, f_H]}|\langle q, a(f,0)\rangle| \leq \gamma,
\end{align}\normalsize
if and only if there exist positive semidefinite \textit{Gram} matrices $\bm{G}_1 \in \mathbb{C}^{n \times n}$ and $\bm{G}_2 \in \mathbb{C}^{(n-1) \times (n-1)}$ such that,\small
\begin{align}
\label{eq:lmi_block1}
\gamma^2 \delta_k = \mathcal{L}_{k, \omega_L, \omega_H} (\bm{G}_1, \bm{G}_2), & \phantom{1} k \in \mathcal{H} \nonumber\\
\begin{bmatrix} \bm{G}_1 & q \\ q^{*} & 1 \end{bmatrix} &\succeq 0,
\end{align}\normalsize
where $\mathcal{H}$ is a halfspace, $\delta_0 = 1$, and $\delta_k = 0$ if $k \neq 0$. This linear matrix inequality representation using positive semidefinite matrix $\bm{G}_1$ paves way for casting the new dual problem in (\ref{eq:weighteddual4}) as a semidefinite program. This above formulation shows that we have changed the inequality form in the convex optimization problem to an equality form allowing semidefinite programming for the weighted atomic norm minimization.

If the cutoff-frequencies $\omega_L$ or $\omega_H$ (in $[-\pi, \pi]$ domain) are equal to $\pm\pi$, then we can write $[\omega_L, \omega_H] = [\omega_L^{'} + \tau, \omega_H^{'} + \tau]$ such that $[\omega_L^{'}, \omega_H^{'}] \subset [-\pi, \pi]$. For the translated subband $[\omega_L^{'}, \omega_H^{'}]$, let the corresponding subband in the domain $[0, 1]$ be $[f_L^{'}, f_H^{'}]$. Then, the LMI formulation given by (\ref{eq:lmi_freq}) becomes valid for this subband. However, the polynomial $q$ is now evaluated in the domain $e^{-i\omega}e^{-i\tau}$ instead of $e^{-i\omega}$. The SDP for this frequency translation employs a scaled version of LMI in (\ref{eq:lmi_block1}),\small
\begin{align}
\delta_k = \mathcal{L}_{k, f_L^{'}, f_H^{'}} (\bm{G}_1, \bm{G}_2), & \phantom{1} k \in \mathcal{H} \nonumber\\
\begin{bmatrix} \bm{G}_1 & \dfrac{1}{\gamma}\tilde{q}_{\tau} \\ \dfrac{1}{\gamma}\tilde{q}^{*}_{\tau} & 1 \end{bmatrix} &\succeq 0,\label{eq:lmi_block3}
\end{align}\normalsize
where\small
\begin{align}
\tilde{q}_{\tau}[j] = q[j]e^{-i\tau j}.&\label{eq:translateddualpoly}
\end{align}\normalsize
We now state the semidefinite program for weighted atomic norm minimization with the probabilistic priors. We use the LMI representation for each of the inequality constraints in (\ref{eq:weighteddual4}) as follows:
\small
\fbox{
 \addtolength{\linewidth}{-2\fboxsep}%
 \addtolength{\linewidth}{-2\fboxrule}%
 \begin{minipage}{\linewidth}
\begin{flalign}
	\label{eq:probpriorSDP}
	\underset{\begin{subarray}{c}
	 q,\\
	 \bm{G}_{11}, \bm{G}_{12}, \cdots, \bm{G}_{1p},\\
	 \bm{G}_{21}, \bm{G}_{22}, \cdots, \bm{G}_{2p}
	\end{subarray}}{\text{maximize}} &\phantom{1} \langle q_{\mathcal{M}},x_{\mathcal{M}} \rangle_{\mathbb{R}}\phantom{1}  \nonumber\\
	\text{subject to   }&\phantom{1} q_{\mathcal{N}\setminus \mathcal{M}}=0\phantom{1}\phantom{1}\phantom{1}\phantom{1}\phantom{1}\phantom{1}\phantom{1}\\
	&\phantom{1} \delta_{k_1} = \mathcal{L}_{k_1, f_{L_1}{'}, f_{H_1}{'}} (\bm{G}_{11}, \bm{G}_{21}),\nonumber\\
	& \phantom{1}\phantom{1}\phantom{1}\phantom{1}\phantom{1}\phantom{1}\phantom{1} k_1 = 0, \cdots, (n-1)\nonumber\\
	& \begin{bmatrix*}[r] \bm{G}_{11} & \dfrac{1}{w_1}\tilde{q}_{\tau_1} \\ \dfrac{1}{w_1}\tilde{q}^*_{\tau_1} & 1 \end{bmatrix*} \succeq 0,\nonumber\\
	&\phantom{1} \delta_{k_2} = \mathcal{L}_{k_2, f_{L_2}{'}, f_{H_2}{'}} (\bm{G}_{12}, \bm{G}_{22}),\nonumber\\
	& \phantom{1}\phantom{1}\phantom{1}\phantom{1}\phantom{1}\phantom{1}\phantom{1} k_2 = 0, \cdots, (n-1)\nonumber\\
	& \begin{bmatrix*}[r] \bm{G}_{12} & \dfrac{1}{w_2}\tilde{q}_{\tau_2} \\ \dfrac{1}{w_2}\tilde{q}^*_{\tau_2} & 1 \end{bmatrix*} \succeq 0,\nonumber\\
	&\phantom{1}\phantom{1}\phantom{1}\phantom{1}\phantom{1}\phantom{1}\phantom{1}\phantom{1}\phantom{1}\phantom{1}\phantom{1}\phantom{1}\vdots\nonumber\\
	&\phantom{1} \delta_{k_p} = \mathcal{L}_{k_p, f_{L_p}{'}, f_{H_p}{'}} (\bm{G}_{1p}, \bm{G}_{2p}),\nonumber\\
	& \phantom{1}\phantom{1}\phantom{1}\phantom{1}\phantom{1}\phantom{1}\phantom{1} k_p = 0, \cdots, (n-1)\nonumber\\
	& \begin{bmatrix*}[r] \bm{G}_{1p} & \dfrac{1}{w_p}\tilde{q}_{\tau_p} \\ \dfrac{1}{w_p}\tilde{q}^*_{\tau_p} & 1 \end{bmatrix*} \succeq 0,\nonumber\\
    \text{where }&\tilde{q}_{\tau_k}[j] = q[j]e^{-i\tau_k j},\phantom{1}k = 1, \cdots, p,\nonumber\\
	&\bm{G}_{11}, \bm{G}_{12}, \cdots, \bm{G}_{1p} \in \mathbb{C}^{n \times n},\nonumber\\
	\text{and }&\bm{G}_{21}, \bm{G}_{22}, \cdots, \bm{G}_{2p} \in \mathbb{C}^{(n-1) \times (n-1)}.\nonumber
\end{flalign}
\end{minipage}
}\normalsize
\vspace{\baselineskip}

The unknown frequencies in $\hat{x}$ can be identified by the frequency localization approach \cite{tang2012csotg} based on computing the dual polynomial, that we state for the weighted atomic norm problem in Algorithm \ref{alg:freqLocal_weighted}. We state that this characterization of the spectral estimation is a general way to integrate given knowledge about the spectrum. If the engineer is able to locate the signal frequency in a particular subband with a very high degree of certainty, better results can be obtained using the optimization (\ref{eq:probpriorSDP}). Also, information about signal frequency bands is frequently available through previous research and measurements, especially in problems pertaining to communication, power systems and remote sensing. We consider this more practical case in the following section.
\begin{algorithm}[t]
\centering
\caption{Frequency localization for probabilistic priors}
\label{alg:freqLocal_weighted}
    \begin{algorithmic}[1]
    \scriptsize
    \STATE Solve the dual problem (\ref{eq:probpriorSDP}) to obtain the optimum solution $q^{\star}$.
	\STATE Let $\mathcal{F} = \{f_1, \cdots, f_j, \cdots, f_s\}$ be the unknown frequencies of signal $x$. The unknown frequencies $f_j$, identify as $\left|\langle q^{\star}, a(f_j, 0) \rangle\right| = w_k$, where $f_j \in \mathcal{B}_k  \subseteq [0, 1]$. For $f \in (\mathcal{B}_k \setminus \mathcal{F}) \subset [0, 1]$, $\left|\langle q^{\star}, a(f, 0) \rangle\right| < w_k$.
	\STATE The corresponding complex coefficients can be recovered by solving a system of simultaneous linear equations $\hat{x}[l] - \sum\limits_{j = 1}^{s}c_ja(f_j, 0)[l] = 0$.
    \end{algorithmic}
\end{algorithm}
\section{Block priors}
\label{sec:block_priors}
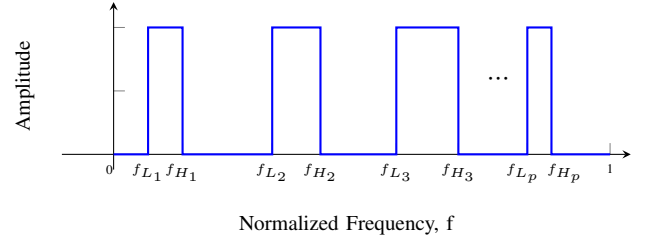
\begin{figure}
\centering
	\begin{tikzpicture}
	\begin{axis}[
		width=3.6in,
		height=1.5in,
		x axis line style={-stealth},
		y axis line style={-stealth},
		xtick={0.5, 1.0, 2.3, 3, 4.1, 5, 6.0, 6.35, 7.2},
		xticklabels={\tiny$f_{L_1}$, \tiny$f_{H_1}$, \tiny$f_{L_2}$, \tiny$f_{H_2}$, \tiny$f_{L_3}$, \tiny$f_{H_3}$, \tiny$f_{L_p}\phantom{1}$, \tiny$\phantom{1}\phantom{1}\phantom{1}f_{H_p}$, \tiny1},
		yticklabels={},
		ymax = 1.2,xmax=7.5,
		axis lines*=center,
		ylabel={\footnotesize Amplitude},
		xlabel={\footnotesize Normalized Frequency, f},
		xlabel near ticks,
		ylabel near ticks]
	\addplot+[thick,mark=none,const plot]
	coordinates
	{(0,0) (0.5,1) (1.0,0) (2.3,1) (3,0) (4.1,1) (5,0) (6.0,1) (6.35,0) (7.2,0)};
	\end{axis}	
	\node [left] at (0.8,  0) {\tiny 0};
	\node [above] at (5.8,  1.0) {$\cdots$};
	\end{tikzpicture}
	\caption{\small The individual frequencies of spectrally sparse signal are assumed to lie in known non-overlapping frequency subbands within the normalized frequency domain $[0, 1]$.}
	\label{fig:mult_bands}
\end{figure}

Of particular interest to spectral estimation are spectrally block sparse signals where certain frequency bands are known to contain all the spectral contents of the signal. Let us assume that all the $s$ frequencies $f_j$ of the spectrally sparse signal $x$ are known \textit{a priori} to lie only in a finite number of non-overlapping frequency bands or intervals within the normalized frequency domain $[0, 1]$. Here, the known set $\mathcal{C}$ is defined as the set $\mathcal{B}$ of all frequency bands in which signal frequencies are known to reside. The prior information consists of the precise locations of all the frequency bands - the lower and upper cut-off frequencies $f_{L_k}$ and $f_{H_k}$ respectively for each of the band $\mathcal{B}_k$ - as shown in the Figure \ref{fig:mult_bands}. We, therefore, have $f_j \in \mathcal{B}, \phantom{1}  \mathcal{B} = \bigcup_{k = 1}^{p} \mathcal{B}_k = \bigcup_{k = 1}^{p} [f_{L_k}, f_{H_k}]$, where $p$ is the total number of disjoint bands known \textit{a priori}.
This \textit{block prior} problem could easily be considered as a special case of probabilistic priors where the probability of a frequency occurring in known subbands is unity while it is zero for all other subbands. When the frequencies are known to reside in the set of subbands $\mathcal{B}$ \textit{a priori}, we propose to minimize a \textit{constrained} atomic norm $||\hat{x}||_{\mathcal{A}, \mathcal{B}}$ for perfect recovery of the signal:\small
\begin{equation}
\label{eq:constrainedatomicnorm}
||\hat{x}||_{\mathcal{A}, \mathcal{B}} = \underset{c_j, f_j \in \mathcal{B}}{\text{inf}}\phantom{1}\left\{\sum\limits_{j=1}^s|c_j|: \hat{x}[l] = \sum\limits_{j=1}^{s} c_je^{i2\pi f_j l} \phantom{1}, \phantom{1} l \in \mathcal{M}\right\}.
\end{equation}\normalsize
As noted earlier, to recover all of the off-the-grid frequencies of the signal $x$ given the block priors, the direct extension of a semidefinite program from (\ref{eq:semiotg}) to minimize the constrained atomic norm is non-trivial. We address this problem by working with the dual problem of the constrained atomic norm minimization, and then transforming the dual problem to an equivalent semidefinite program by using theories of positive trigonometric polynomials. We note that in the case of block priors, (\ref{eq:dualatomicnorm}) can be written as $\|q\|_{\mathcal{A}, \mathcal{B}}^* = \sup_{f \in \mathcal{B}}|\langle q, a(f,0)\rangle| = \sup_{f \in \mathcal{B}}|Q(f)|$, where $Q(f)$ is the dual polynomial. The primal problem of constrained atomic norm minimization is given by\small
\begin{flalign}
	\label{eq:blockpriorprimal}
	& \underset{x}{\text{minimize}}\phantom{1}  \|x\|_{\mathcal{A}, \mathcal{B}}\nonumber\\
	& \text{subject to}\phantom{1} \hat{x}[l] = x[l], \phantom{1} l \in \mathcal{M},
\end{flalign}\normalsize
and, similar to (\ref{eq:dualtoatomicminimization}), we can formulate the corresponding dual problem as\small
\begin{flalign}
	\label{eq:dual_withblockpriors}
	 \underset{q}{\text{maximize}} & \phantom{1} \langle q_{\mathcal{M}},x_{\mathcal{M}}\rangle_{\mathbb{R}}\phantom{1}  \nonumber\\
	 \text{subject to}&\phantom{1} q_{\mathcal{N}\setminus \mathcal{M}}=0\\
	 &\phantom{1} \|q\|_{\mathcal{A}, \mathcal{B}}^* \leq 1, \nonumber
\end{flalign}\normalsize
where $\|q\|_{\mathcal{A}, \mathcal{B}}^* = \sup_{f \in \mathcal{B}}|\langle q, a(f,0)\rangle|$. Since $\mathcal{B}$ is defined as a union of multiple frequency bands, the inequality constraint in (\ref{eq:dual_withblockpriors}) can be expanded to $p$ separate inequality constraints. It can be easily observed that (\ref{eq:dual_withblockpriors}) is a special case of (\ref{eq:weighteddual2}) with all the weights being unity and $\mathcal{B} \subseteq [0, 1]$ (i. e. the set of bands $\mathcal{B}$ need not necessarily cover the entire frequency range).
While framing the semidefinite program for this problem, we use a linear matrix inequality similar to that in (\ref{eq:lmi_block1}) with $\gamma = 1$ for each of the inequality constraint in (\ref{eq:dual_withblockpriors}), to cast the dual problem constraint into a semidefinite program. So, when all the frequencies are known to lie in $p$ disjoint frequency bands, then the semidefinite program for the dual problem in (\ref{eq:dual_withblockpriors}) can be constructed by using $p$ equality-form constraints:
\small
\fbox{
 \addtolength{\linewidth}{-2\fboxsep}%
 \addtolength{\linewidth}{-2\fboxrule}%
 \begin{minipage}{\linewidth}
\begin{flalign}
	\label{eq:blocksparsitySDP}
	\underset{\begin{subarray}{c}
	 q,\\
	 \bm{G}_{11}, \bm{G}_{12}, \cdots, \bm{G}_{1p},\\
	 \bm{G}_{21}, \bm{G}_{22}, \cdots, \bm{G}_{2p}
	\end{subarray}}{\text{maximize}} &\phantom{1} \langle q_{\mathcal{M}},x_{\mathcal{M}} \rangle_{\mathbb{R}}\phantom{1}  \nonumber\\
	\text{subject to   }&\phantom{1} q_{\mathcal{N}\setminus \mathcal{M}}=0\phantom{1}\phantom{1}\phantom{1}\phantom{1}\phantom{1}\phantom{1}\phantom{1}\\
	&\phantom{1} \delta_{k_1} = \mathcal{L}_{k_1, f_{L_1}, f_{H_1}} (\bm{G}_{11}, \bm{G}_{21}),\nonumber\\
	& \phantom{1}\phantom{1}\phantom{1}\phantom{1}\phantom{1}\phantom{1}\phantom{1} k_1 = 0, \cdots, (n-1)\nonumber\\
	& \begin{bmatrix*}[r] \bm{G}_{11} & q \\ q^* & 1 \end{bmatrix*} \succeq 0,\nonumber\\
	&\phantom{1} \delta_{k_2} = \mathcal{L}_{k_2, f_{L_2}, f_{H_2}} (\bm{G}_{12}, \bm{G}_{22}),\nonumber\\
	& \phantom{1}\phantom{1}\phantom{1}\phantom{1}\phantom{1}\phantom{1}\phantom{1} k_2 = 0, \cdots, (n-1)\nonumber\\
	& \begin{bmatrix*}[r] \bm{G}_{12} & q \\ q^* & 1 \end{bmatrix*} \succeq 0,\nonumber\\
	&\phantom{1}\phantom{1}\phantom{1}\phantom{1}\phantom{1}\phantom{1}\phantom{1}\phantom{1}\phantom{1}\phantom{1}\phantom{1}\phantom{1}\vdots\nonumber\\
	&\phantom{1} \delta_{k_p} = \mathcal{L}_{k_p, f_{L_p}, f_{H_p}} (\bm{G}_{1p}, \bm{G}_{2p}),\nonumber\\
	& \phantom{1}\phantom{1}\phantom{1}\phantom{1}\phantom{1}\phantom{1}\phantom{1} k_p = 0, \cdots, (n-1)\nonumber\\
	& \begin{bmatrix*}[r] \bm{G}_{1p} & q \\ q^* & 1 \end{bmatrix*} \succeq 0,\nonumber\\
	\text{where }&\bm{G}_{11}, \bm{G}_{12}, \cdots, \bm{G}_{1p} \in \mathbb{C}^{n \times n},\nonumber\\
	\text{and }&\bm{G}_{21}, \bm{G}_{22}, \cdots, \bm{G}_{2p} \in \mathbb{C}^{(n-1) \times (n-1)}.\nonumber
\end{flalign}
\end{minipage}
}\normalsize
\vspace{\baselineskip}

In the extreme case when any of the known frequency bands $\mathcal{B}_k$ have $\omega_{L_k}$ or $\omega_{H_k}$ lying exactly on either $-\pi$ or $\pi$, then the dual-polynomial in \ref{eq:blocksparsitySDP} should be appropriately translated as noted in (\ref{eq:translateddualpoly}).

In many applications, the location of \textit{some} of the signal frequencies might be precisely known. One could think of this \textit{known poles} problem as a probabilistic prior problem where the cardinality of \textit{some} sets $\mathcal{B}_k$ is exactly unity (and the associated probability be unity as well), while the remaining frequency subbands have a non-unity probability. However, there are a few differences. For probabilistic priors, the probability distribution function is known for the entire interval $[0, 1]$ while, in case of known poles, the probability distribution of the bands of unknown frequencies is unavailable. Also, unlike block prior formulation, known poles problem does not have zero probability associated with the remaining subbands.
\section{Known Poles}
\label{sec:known_poles}
We now consider the case when some frequency components are known \emph{a priori} but their corresponding amplitudes and phases are not. Let the index set of all the frequencies be $\mathcal{S}$, $|\mathcal{S}| = s$. Let $\mathcal{P}$ be the index set of all the known frequencies, and $|\mathcal{P}| = p$. Namely, we assume that the signal $x$ contains some known frequencies $f_j$, $j \in \mathcal{P} \subseteq \mathcal{S}$, $|\mathcal{P}| = p$. For known frequencies, let us denote their \emph{complex} coefficients as $d_j$ and their \textit{phaseless frequency atoms} as $a_{j}[l] = a(f_j, 0)[l]  = e^{i2\pi f_j l}$. We define the \textit{conditional atomic norm} $||\hat{x}||_{\mathcal{A},\mathcal{P}}$ for the \textit{known poles} as follows:\\\small
\begin{flalign}
    \label{eq:condatomicnorm}
    & ||\hat{x}||_{\mathcal{A},\mathcal{P}} = \underset{c_j, d_j, f_j}{\text{inf}}\phantom{1}\left\{\sum\limits_{j=1}^{s-p}|c_j|: \hat{x}[l] = \sum\limits_{j=1}^{s-p} c_je^{i2\pi f_jl} \right.\nonumber\\
    & \phantom{1}\phantom{1}\phantom{1}\phantom{1}\phantom{1}\phantom{1}\phantom{1}\phantom{1}\phantom{1}\phantom{1}\phantom{1}\phantom{1}\phantom{1}\phantom{1}\phantom{1}\phantom{1}\phantom{1}\left.+ \sum\limits_{j=s-p+1}^{s} d_je^{i2\pi f_jl}\phantom{1}, \phantom{1} l \in \mathcal{M}\right\}.
\end{flalign}\normalsize
Unlike previously mentioned \textit{a priori} counterparts of the atomic norm, the semidefinite formulation for $||\hat{x}||_{\mathcal{A},\mathcal{P}}$ easily follows from (\ref{eq:atomicnorm_sdpdef}).
\begin{proposition}
The conditional atomic norm for a vector $\hat{x}$ is given by\small
\begin{equation}
\label{eq:condatomicnorm_semidef}
||\hat{x}||_{\mathcal{A},\mathcal{P}} = \underset{T_n, \tilde{x}, t, d_j}{\text{inf}}\left\{\dfrac{1}{2|\mathcal{N}|} \text{Tr($T_n$)} + \frac{1}{2}t : \begin{bmatrix*}[r] T_n & \tilde{x} \\ \tilde{x}^* & t \end{bmatrix*} \succeq 0 \right\},
\end{equation}\normalsize
where $\tilde{x}[l] = \hat{x}[l] - \sum\limits_{j \in \mathcal{P}}a_j[l]d_j$ represents the positive combination of complex sinusoids with unknown poles.
\label{prop:cond_atomicnorm}
\end{proposition}
\begin{IEEEproof}[Proof of Proposition \ref{prop:cond_atomicnorm}.] From (\ref{eq:condatomicnorm}), we simply have $\tilde{x}[l] = \hat{x}[l] - \sum\limits_{j \in \mathcal{P}}a_j[l]d_j = \sum\limits_{j=1}^{s-p} c_je^{i2\pi f_jl}$, meaning the value of the semidefinite program in (\ref{eq:condatomicnorm_semidef}) is same as $||\tilde{x}||_{\mathcal{A}} = ||\hat{x}||_{\mathcal{A},\mathcal{P}}$.
\end{IEEEproof}
The conditional atomic norm minimization problem can be posed as the following semidefinite formulation in a similar way as in (\ref{eq:semiotg}):\small
\begin{flalign}
\label{eq:semiprior}
	& \underset{T_n, \hat{x}, \tilde{x}, t, d_j}{\text{minimize}}\phantom{1} \dfrac{1}{2|\mathcal{N}|}  \text{Tr($T_n$)} + \frac{1}{2}t\nonumber\\
	& \text{subject to}\phantom{1} \begin{bmatrix*}[r] T_n & \tilde{x} \\ \tilde{x}^* & t \end{bmatrix*} \succeq 0\\
	& \hat{x}[l] = x[l], \phantom{1} l \in \mathcal{M}\nonumber\\
	& \hat{x}[l] = \tilde{x}[l] + \sum\limits_{j \in \mathcal{P}}a_j[l]d_j, \phantom{1} l \in \mathcal{M}.\nonumber
\end{flalign}\normalsize
$\tilde{x}$ can be viewed as the signal filtered of the \textit{known poles}. The remaining unknown frequencies in $\tilde{x}$ can be identified by the frequency localization approach that we restate for $\tilde{x}$ in Algorithm \ref{alg:freqLocal}.
\begin{algorithm}[t]
\centering
\caption{\textit{Known poles} algorithm}
\label{alg:freqLocal}
    \begin{algorithmic}[1]
    \scriptsize
    \STATE Solve the semidefinite program (\ref{eq:semiprior}) to obtain $\tilde{x}$.
    \STATE Solve the following dual problem to obtain the optimum solution $q^{\star}$
	\begin{flalign}
		& \underset{q}{\text{maximize}}\phantom{1} \langle q, \tilde{x}\rangle_{\mathbb{R}}\nonumber\\
		& \text{subject to}\phantom{1} ||q||^*_\mathcal{A} \le 1\\
		& q[l] = 0, \phantom{1} l \in \mathcal{N} \setminus \mathcal{M}.\nonumber
	\end{flalign}
	\STATE The unknown frequencies $f_j$, $j \in \mathcal{P}$, identify as $\left|\langle q^{\star}, a_j \rangle\right| = 1$. For $j \notin \mathcal{S} \setminus \mathcal{P}$, $\left|\langle q^{\star}, a_j \rangle\right| < 1$.
	\STATE Solve the following system of simultaneous linear equations to recover the complex coefficients of unknown frequencies: $\tilde{x}[l] - \sum\limits_{j \in \mathcal{S} \setminus \mathcal{P}}c_ja_j[l] = 0$.
    \end{algorithmic}
\end{algorithm}

\section{Performance Analysis}
\label{sec:perf_anal}
To identify the true frequencies of the signal from the solution of the dual problem, we now establish the conditions for finding the dual-certificate of support when prior information is available. We additionally show that the dual polynomial requirements can be slackened if the prior information gives the approximate location of each of the signal frequencies. We further put our result in the context of minimum number of signal samples required for the reconstruction of the signal $x$.

Since the primal problem (\ref{eq:primalweighted}) has only equality constraints, Slater's condition is satisfied. As a consequence, strong duality holds \cite{boyd2004convex}. This allows us to present the dual-certificate of support for the optimizer of (\ref{eq:primalweighted}). In the following theorems, $\text{sign}(c_j) = \nicefrac{c_j}{|c_j|}$, and $\Re(\cdot)$ denotes the real part (of a complex number).
\begin{theorem} Let the set of atoms $\{a_{\mathcal{M}}(f_1, 0), \cdots, a_{\mathcal{M}}(f_s, 0)\}$ supported on subset $\mathcal{M}$ of $\mathcal{N}$ be linearly independent. Then, $\hat{x} = x$ is the unique solution to the primal problem (\ref{eq:primalweighted}), if there exists a polynomial\small
\begin{align}
\label{eq:dualpoly_dc}
Q(f) = \langle q, a(f,0) \rangle = \sum\limits_{l=0}^{n-1} q_{l}e^{-i 2\pi f l },
\end{align}\normalsize
such that\small
\begin{align}
Q(f_j) &= w_k\text{sign}(c_j), \forall f_j \in \mathcal{B}_k \subseteq [0, 1]\label{eq:condition1_thm_dualcert}\\
|Q(f)| &< w_k, \forall f \in (\mathcal{B}_k \setminus \mathcal{F}) \subset [0, 1]\label{eq:condition2_thm_dualcert}\\
q_{\mathcal{N}\setminus \mathcal{M}}&=0.\label{eq:condition3_thm_dualcert}
\end{align}\normalsize
\label{thm:dual_cert_blockprior}
\end{theorem}
\begin{IEEEproof}[Proof of Theorem \ref{thm:dual_cert_blockprior}.] The proof follows from the dual polynomial for the standard atomic norm minimization problem. We refer the reader to \cite{tang2012csotg} for details. Briefly, it can be concluded that strong duality holds and we have $\langle q_{\mathcal{M}}, x_{\mathcal{M}}\rangle_{\mathbb{R}} = \sum\limits_{j=1}^{s}w_j|c_j| = ||x||_{\mathbf{w}\mathcal{A}}$, where the vector $q$ satisfies the conditions (\ref{eq:condition1_thm_dualcert}), (\ref{eq:condition2_thm_dualcert}), and (\ref{eq:condition3_thm_dualcert}), and is dual feasible. As for the uniqueness, let $x^{\dagger}[l] = \sum\limits_{j}c_j^{\dagger}e^{i2\pi f_j^{\dagger}l}$, $l \in \mathcal{M}$, be an alternative minimizer of (\ref{eq:primalweighted}) such that $x^{\dagger}$ contains frequencies outside the set $\mathcal{F}$ of oracle frequencies. Then,\small
\begin{align}
\begin{aligned}
&||x||_{\mathbf{w}\mathcal{A}} = \langle q_{\mathcal{M}}, x_{\mathcal{M}}\rangle_{\mathbb{R}} = \langle q_{\mathcal{M}}, x^{\dagger}_{\mathcal{M}}\rangle_{\mathbb{R}} = \langle q_{\mathcal{M}}, \sum\limits_{j}c_j^{\dagger}e^{i2\pi f_j^{\dagger}l}\rangle_{\mathbb{R}} \nonumber\\
&= \left\langle q_{\mathcal{M}}, \sum\limits_k\sum\limits_{f_j \in \mathcal{F} \subset \mathcal{B}_k}c_j^{\dagger}e^{i2\pi f_jl} + \sum\limits_k\sum\limits_{f^{\dagger}_{h} \in (\mathcal{B}_k\setminus\mathcal{F})}c^{\dagger}_{h}e^{i2\pi f^{\dagger}_{h}l}\right\rangle_{\mathbb{R}} \\
&< \sum\limits_{f_j \in \mathcal{F}\subset \mathcal{B}_k}w_j|c_j^{\dagger}| + \sum\limits_{f^{\dagger}_{h} \in (\mathcal{B}_k\setminus\mathcal{F})}w_h|c^{\dagger}_{h}|\\
&= ||x^{\dagger}||_{\mathbf{w}\mathcal{A}},
\end{aligned}
\end{align}\normalsize
resulting in a contradiction that $x^{\dagger}$ is not a minimizer of (\ref{eq:primalweighted}). If $x^{\dagger}$ contains only the oracle frequencies and the same sign pattern $\nicefrac{c_j}{|c_j|}$ as that of $x$, then $x^{\dagger}$ also has the same complex coefficients as $x$ since the set $\{a_{\mathcal{M}}(f_1, 0), \cdots, a_{\mathcal{M}}(f_s, 0)\}$ is linearly independent. Therefore, the optimal solution is unique.
\end{IEEEproof}
As a corollary to Theorem \ref{thm:dual_cert_blockprior}, we can arrive at the dual polynomial for the block prior problem as follows.
\begin{corollary}
The $\hat{x} = x$ is the unique solution to the primal problem (\ref{eq:blockpriorprimal}), if there exists a polynomial $Q(f)$
such that\small
\begin{align}
Q(f_j) &= \text{sign}(c_j), \forall f_j \in \mathcal{F} \subset \mathcal{B}\label{eq:condition1_thm_dualcert2}\\
|Q(f)| &< 1, \forall f \in (\mathcal{B}\setminus\mathcal{F})\label{eq:condition2_thm_dualcert2}\\
q_{\mathcal{N}\setminus \mathcal{M}}&=0.\label{eq:condition3_thm_dualcert2}
\end{align}\normalsize
\end{corollary}
When the prior information is available to such a generous extent that each of the individual frequencies are known within close boundaries, as we present next, an interesting consequence of this relaxation is that the number of samples required to reconstruct the signal could be bounded.
\begin{theorem} Let the signal $x$ as in (\ref{eq:sigmodelstd}) be sampled on a subset $\mathcal{M}$ of $\mathcal{N}$. If there exists a polynomial $Q(f)$ such that $\forall f_j \in \mathcal{F} \subset \mathcal{B}$,\small
\begin{align}
	Q(f_j) &= \text{sign}(c_j)\label{eq:signcondition_alternate}\\
    Q^{'}(f_j) &= \sum\limits_{l=0}^{n-1} l q_{l}e^{-i 2\pi f_j l } = 0\label{eq:firstderivcondition_alternate}\\
	Q^{''}(f_j) &= \sum\limits_{l=0}^{n-1} - (2 \pi l)^2 q_{l}e^{-i 2\pi f_j l } = -\text{sign}(\Re(c_j)),\label{eq:secondderivcondition_alternate}	
\end{align}\normalsize
and, if each of the frequencies is known within a sufficiently small frequency subband, then $\hat{x} = x$ is the unique optimizer of the primal problem (\ref{eq:blockpriorprimal}). Further, assuming $f_j$s are distributed uniformly at random in $[0,1]$, such a dual polynomial exists with probability $1$ when $m \geq 3s$.
\label{thm:theorem_3s}
\end{theorem}
\begin{IEEEproof}[Proof of Theorem \ref{thm:theorem_3s}.] The polynomial that we seek can be written as $Q(f) = Q_R(f) + iQ_I(f)$, where $Q_R(f)$ and $Q_I(f)$ are the real and imaginary parts respectively. As per Theorem \ref{thm:dual_cert_blockprior}, $Q(f)$ should also satisfy the conditions (\ref{eq:condition1_thm_dualcert}) and (\ref{eq:condition2_thm_dualcert}). Therefore, (\ref{eq:signcondition_alternate}) is a restatement of (\ref{eq:condition1_thm_dualcert}) as follows:\small
\begin{align}
\label{eq:signcondition}
	Q(f_j) = \sum\limits_{l=0}^{n-1} q_{l}e^{-i 2\pi f_j l } = \text{sign}(c_j) = \frac{c_j}{|c_j|}\phantom{1}\forall f_j \in \mathcal{F} \subset \mathcal{B}.
\end{align}\normalsize
For the dual polynomial to achieve an extremum at $f_j \in \mathcal{F} \subset \mathcal{B}$ as specified by (\ref{eq:condition2_thm_dualcert}), the following is a sufficient condition for its first derivative leading to (\ref{eq:firstderivcondition_alternate}):\small
\begin{align}
\label{eq:firstderivcondition}
	Q^{'}(f_j) = \sum\limits_{l=0}^{n-1} -i 2\pi l q_{l}e^{-i 2\pi f_j l } = 0\phantom{1}\forall f_j \in \mathcal{F} \subset \mathcal{B}.
\end{align}\normalsize
The condition for a maximum at $f_j \in \mathcal{F} \subset \mathcal{B}$ requires the second derivative $|Q(f_j)|^{''}$ to be strictly negative. We have,\small
\begin{align}
\label{eq:secondderivative}
	|Q(f_j)|^{''} &= -\frac{[Q_R(f_j)Q_R^{'}(f_j) + Q_I(f_j)Q_I^{'}(f_j)]^2}{|Q(f_j)|^3}\nonumber\\
			 &+ \frac{|Q^{'}(f_j)|^2 + Q_R(f_j)Q_R^{''}(f_j) + Q_I(f_j)Q_I^{''}(f_j)}{|Q(f_j)|}\nonumber\\
			 &\phantom{1}\forall f_j \in \mathcal{F} \subset \mathcal{B}.
\end{align}\normalsize
Therefore, for $|Q(f_j)|^{''}$ to be strictly negative, it is sufficient to require,\small
\begin{align}
\label{eq:secondderivcondition}
	|Q^{'}(f_j)|^2 + Q_R(f_j)Q_R^{''}(f_j) + Q_I(f_j)Q_I^{''}(f_j) < 0\phantom{1}\forall f_j \in \mathcal{F} \subset \mathcal{B}.
\end{align}\normalsize
Under the condition (\ref{eq:secondderivcondition}), when the frequencies $f_j$ are known to lie in a very small frequency band $\mathcal{B}_k$ such that $(f_{H_k} - f_{L_k}) \ll 1$, then the polynomial constraints are valid within such a sufficiently small interval. 

To satisfy the constraint (\ref{eq:secondderivcondition}), we impose an additional constraint that requires $Q_I^{''}(f_j)$ to vanish, reducing (\ref{eq:secondderivcondition}) to
\begin{align}
\label{eq:secondderivcondition_real}
	Q_R(f_j)Q_R^{''}(f_j) < 0\phantom{1}\forall f_j \in \mathcal{F} \subset \mathcal{B}.
\end{align}
Using the definition of dual polynomial from (\ref{eq:dualpoly_dc}), we can now cast (\ref{eq:secondderivcondition_real}) as the condition (\ref{eq:secondderivcondition_alternate}).

Let $x_{f_j} = e^{i2 \pi f_j}$. We show that the linear system (\ref{eq:signcondition_alternate}), (\ref{eq:firstderivcondition_alternate}), and (\ref{eq:secondderivcondition_alternate}) results in a unique solution, given at least $3s$ equations as follows: 
\begin{align}
	\underbrace{\begin{bsmallmatrix}	
    	x_{f_1}^{l_0} & x_{f_1}^{l_1} & \cdots & x_{f_1}^{l_{3s-1}}\\
		l_0 x_{f_1}^{l_0} & l_1 x_{f_1}^{l_1} & \cdots & l_{3s-1} x_{f_1}^{l_{3s-1}}\\[5pt]
		-(2\pi l_0)^2 x_{f_1}^{l_0} & -(2\pi l_1)^2 x_{f_1}^{l_1} & \cdots & -(2\pi l_{3s-1})^2 x_{f_1}^{l_{3s-1}}\\[5pt]
		x_{f_2}^{l_0} & x_{f_2}^{l_1} & \cdots & x_{f_2}^{l_{3s-1}}\\[5pt]
		l_0 x_{f_2}^{l_0} & l_1 x_{f_2}^{l_1} & \cdots & l_{3s-1} x_{f_2}^{l_{3s-1}}\\[5pt]
		-(2\pi l_0)^2 x_{f_2}^{l_0} & -(2\pi l_1)^2 x_{f_2}^{l_1} & \cdots & -(2\pi l_{3s-1})^2 x_{f_2}^{l_{3s-1}}\\[5pt]
        \vdots & \vdots & \ddots & \vdots \\
		x_{f_s}^{l_0} & x_{f_s}^{l_1} & \cdots & x_{f_s}^{l_{3s-1}}\\[5pt]
		l_0 x_{f_s}^{l_0} & l_1 x_{f_s}^{l_1} & \cdots & l_{3s-1} x_{f_s}^{l_{3s-1}}\\[5pt]
		-(2\pi l_0)^2 x_{f_s}^{l_0} & -(2\pi l_1)^2 x_{f_s}^{l_1} & \cdots & -(2\pi l_{3s-1})^2 x_{f_s}^{l_{3s-1}}
	\end{bsmallmatrix}}_\text{$= \bm{A}$}
    \begin{bsmallmatrix} q_{l_0} \\[9pt] q_{l_0} \\[9pt] q_{l_0} \\[9pt] q_{l_1} \\[9pt] q_{l_1} \\[9pt] q_{l_1} \\[9pt] \vdots \\[9pt] q_{l_{3s-1}} \\[9pt] q_{l_{3s-1}} \\[9pt] q_{l_{3s-1}} \end{bsmallmatrix}\nonumber\\
    = \begin{bsmallmatrix} \frac{c_1}{|c_1|} & 0 & -\text{sign}(\Re(c_1)) & \frac{c_2}{|c_2|} & 0 & -\text{sign}(\Re(c_2)) & \cdots & \frac{c_s}{|c_s|} & 0 & -\text{sign}(\Re(c_s))\end{bsmallmatrix}^T,
\label{eq:lin_eqns_3s}
\end{align}
where $l_0, l_1, \cdots, l_{3s-1}$ are the indices of the samples of the signal $x$. Proposition \ref{prop:fullrank_3s} completes the proof by showing that the system matrix $\bm{A}$ in (\ref{eq:lin_eqns_3s}) is invertible with probability 1, provided the frequencies in the set $\mathcal{F} = \{f_1, \cdots, f_j, \cdots, f_s\}$ are distributed uniformly at random.
\end{IEEEproof}
\begin{proposition}
Let $\mathcal{M} = \{l_0, l_1, \cdots, l_{3s-1}\}$ be the set of indices for $3s$ samples of the signal $x$. Let $h_{f_j} = e^{i2 \pi f_j}$, then the $3s \times 3s$ matrix\small
\begin{align}
	\bm{A}_s = \begin{bmatrix}	
    	h_{f_1}^{l_0} & h_{f_1}^{l_1} & \cdots & h_{f_1}^{l_{3s-1}}\\[5pt]
		l_0 \cdot h_{f_1}^{l_0} & l_1 \cdot h_{f_1}^{l_1} & \cdots & l_{3s-1} \cdot h_{f_1}^{l_{3s-1}}\\[5pt]
		l_0^2 \cdot h_{f_1}^{l_0} & l_1^2 \cdot h_{f_1}^{l_1} & \cdots & l_{3s-1}^2 \cdot h_{f_1}^{l_{3s-1}}\\[5pt]
		h_{f_2}^{l_0} & h_{f_2}^{l_1} & \cdots & h_{f_2}^{l_{3s-1}}\\[5pt]
		l_0 \cdot h_{f_2}^{l_0} & l_1 \cdot h_{f_2}^{l_1} & \cdots & l_{3s-1} \cdot h_{f_2}^{l_{3s-1}}\\[5pt]
		l_0^2 \cdot h_{f_2}^{l_0} & l_1^2 \cdot h_{f_2}^{l_1} & \cdots & l_{3s-1}^2 \cdot h_{f_2}^{l_{3s-1}}\\[5pt]
        \vdots & \vdots & \ddots & \vdots \\[5pt]
		h_{f_s}^{l_0} & h_{f_s}^{l_1} & \cdots & h_{f_s}^{l_{3s-1}}\\[5pt]
		l_0 \cdot h_{f_s}^{l_0} & l_1 \cdot h_{f_s}^{l_1} & \cdots & l_{3s-1} \cdot h_{f_s}^{l_{3s-1}}\\[5pt]
		l_0^2 \cdot h_{f_s}^{l_0} & l_1^2 \cdot h_{f_s}^{l_1} & \cdots & l_{3s-1}^2 \cdot h_{f_s}^{l_{3s-1}}
	\end{bmatrix},
\label{eq:matrix_fullrank}    
\end{align}\normalsize
is full rank with probability 1 if the frequencies $f_1, \cdots, f_j, \cdots, f_s$ are drawn uniformly at random in $[0, 1]$.
\label{prop:fullrank_3s}
\end{proposition}
\begin{IEEEproof}[Proof of Proposition \ref{prop:fullrank_3s}.]
We show $\bm{A}_s$ is full-rank by proving that its determinant, $det(\bm{A}_s) = |\bm{A}_s|$ is a non-zero polynomial. For $s = 1$, we have the matrix,\small
\begin{align}
\bm{A}_1 = \begin{bmatrix}	
    	h_{f_1}^{l_0} & h_{f_1}^{l_1} & h_{f_1}^{l_2}\\[5pt]
		l_0 \cdot h_{f_1}^{l_0} & l_1 \cdot h_{f_1}^{l_1} & l_2 \cdot h_{f_1}^{l_2}\\[5pt]
		l_0^2 \cdot h_{f_1}^{l_0} & l_1^2 \cdot h_{f_1}^{l_1} & l_2^2 \cdot h_{f_1}^{l_2}
	\end{bmatrix}.
\label{eq:s_equals_1}    
\end{align}\normalsize
We note that $|\bm{A}_s|$ easily reduces to a Vandermonde determinant (of order $3$), so that $|\bm{A}_s| = (l_2 - l_1)(l_2 - l_0)(l_1 - l_0)h_{f_1}^{l_0 + l_1 + l_2}$, which is a non-zero polynomial because $l_0$, $l_1$, and $l_2$ are distinct sample indices.

Let us now assume that, for $s>1$, $|\bm{A}_s|$ is a non-zero polynomial. We would like to show that $|\bm{A}_{s+1}|$ is also a non-zero polynomial. We have,\small
\begin{align}
\bm{A}_{s+1} &= \begin{bmatrix} \bm{A}_{s} & \bm{B} \\ \bm{C} & \bm{D}\end{bmatrix},
\end{align}\normalsize
where 
\begin{align}\small 
\bm{D} &= \begin{bmatrix}	
    	h_{f_{s+1}}^{l_{3s}} & h_{f_{s+1}}^{l_{3s+1}} & h_{f_{s+1}}^{l_{3s+2}}\\[5pt]
		l_{3s} h_{f_{s+1}}^{l_{3s}} & l_{3s+1} h_{f_{s+1}}^{l_{3s+1}} & l_{3s+2} h_{f_{s+1}}^{l_{3s+2}}\\[5pt]
		l_{3s}^2 h_{f_{s+1}}^{l_{3s}} & l_{3s+1}^2 h_{f_{s+1}}^{l_{3s+1}} & l_{3s+2}^2 h_{f_{s+1}}^{l_{3s+2}}
	\end{bmatrix}_{3 \times 3}.
\end{align}\normalsize
Noting that the determinant of row echelon form is same as the original matrix, we obtain the row echelon form (REF) for the matrix $\bm{D}$ as follows:
\begin{align}\small
REF(\bm{D}) = \begin{bsmallmatrix}	
    	h_{f_{s+1}}^{l_{3s}} & h_{f_{s+1}}^{l_{3s+1}} & h_{f_{s+1}}^{l_{3s+2}}\\[5pt]
		0 & (l_{3s+1} - l_{3s})h_{f_{s+1}}^{l_{3s+1}} & (l_{3s+2} - l_{3s}) h_{f_{s+1}}^{l_{3s+2}}\\[5pt]
		0 & 0 & (l_{3s+2} - l_{3s})(l_{3s+2} - l_{3s+1}) h_{f_{s+1}}^{l_{3s+2}}
	\end{bsmallmatrix}.
\label{eq:rref1}
\end{align}\normalsize
Let $a_{i,j}$ be the element of the matrix $\bm{A}_{s+1}$ in $i$th row and $j$th column, then by the Leibniz formula for determinants,\small
\begin{align}
|\bm{A}_{s+1}| &= \sum_{\sigma \in S_{3s+3}}sgn(\sigma)a_{1,\sigma(1)}a_{2,\sigma(2)}...a_{3s+3,\sigma(3s+3)}\label{eq:leib_asp1}\\
&= P_{l_{3s}+l_{3s+1}+l_{3s+2}}(h_{f_{s+1}})|\bm{A}_s| + P(h_{f_1}, \cdots, h_{f_s}, h_{f_{s+1}}),\nonumber
\end{align}\normalsize
where $\text{sgn}$ is the sign function of permutations in the permutation group $S_{s+1}$, $P_{l_{3s}+l_{3s+1}+l_{3s+2}}(h_{f_{s+1}})$ is a non-zero univariate monomial in $h_{f_{s+1}}$ of degree ${l_{3s}+l_{3s+1}+l_{3s+2}}$, and $P(h_{f_1}, \cdots, h_{f_s}, h_{f_{s+1}})$ is a multivariate polynomial. From the row echelon form in (\ref{eq:rref1}), we recognize that the highest degree of the variable $h_{f_{s+1}}$ in the expansion (\ref{eq:leib_asp1}) is $l_{3s}+l_{3s+1}+l_{3s+2}$. Note that the polynomial $P(h_{f_1}, \cdots, h_{f_s}, h_{f_{s+1}})$ has lower degree in $h_{f_{s+1}}$ than $P_{l_{3s}+l_{3s+1}+l_{3s+2}}(h_{f_{s+1}})$. Since $|\bm{A}_s|$ is a non-zero polynomial, the coefficient of $h_{f_{s+1}}^{l_{3s}+l_{3s+1}+l_{3s+2}}$ is also a non-zero polynomial. Therefore, $|\bm{A}_{s+1}|$ is a non-zero polynomial. Further, the probability that one randomly picks the frequencies over $[0, 1]$ such that each $h_{f_j}$ is a root of this non-zero polynomial is zero.\footnote{An analogous argument for a polynomial with roots over a finite field can be found in Schwartz-Zippel-DeMillo-Lipton lemma \cite{schwartz1979probabilistic, zippel1979probabilistic, demillo1978probabilistic}.} Thus, by induction, $|\bm{A}_s|$ is non-zero with probability 1.
\end{IEEEproof}
The formulation in (\ref{eq:probpriorSDP}) generalizes the prior information. As the cases of \textit{block priors} and \textit{known poles} indicate, the more we know about the spectral structure of the signal, precise formulations of atomic norm minimization can be evaluated to boost signal recovery. If all poles are known in the sense of \textit{known poles} algorithm (i.e., the amplitudes and phases of all known poles are unknown), then the signal $x$ can be uniquely reconstructed using the randomly sampled support $x_{\mathcal{M}}$ where $|\mathcal{M}| = s$. Further, it is well known that if the signal is uniformly sampled, then the Prony's method can uniquely reconstruct the signal $x$ using no more than $2s$ samples. In comparison, our results from Theorem \ref{thm:theorem_3s} show that if each of the poles are \textit{approximately known}, then the unique reconstruction of the signal $x$ requires no more than $3s$ samples.

\section{Numerical Experiments}
\label{sec:numsim}
\begin{figure}[!t]
\centering
\subfloat[Without any priors]{%
  \includegraphics[width=3.5cm]{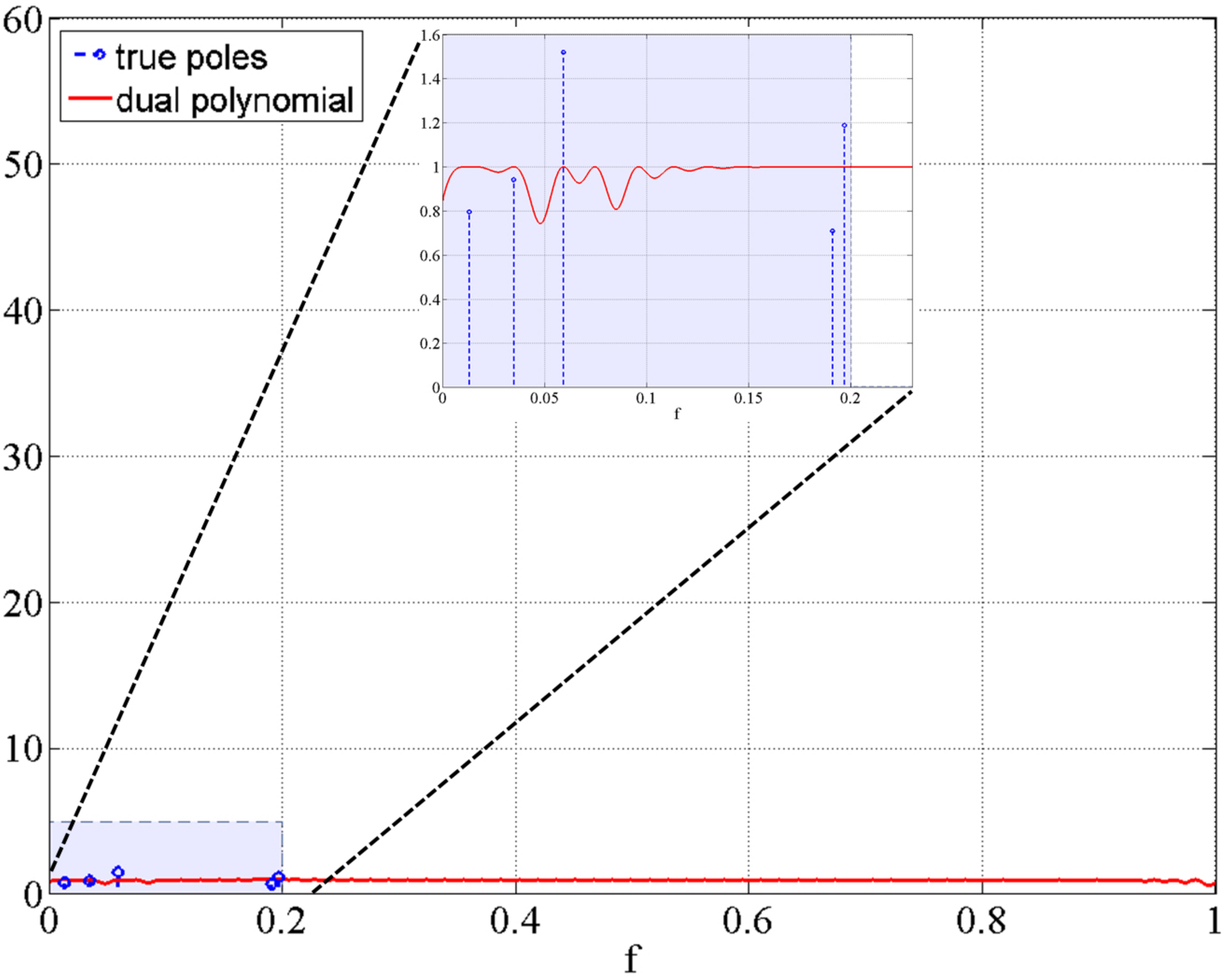}%
  \label{fig:no_probpriors}%
}\qquad
\subfloat[With probabilistic priors]{%
  \includegraphics[width=3.5cm]{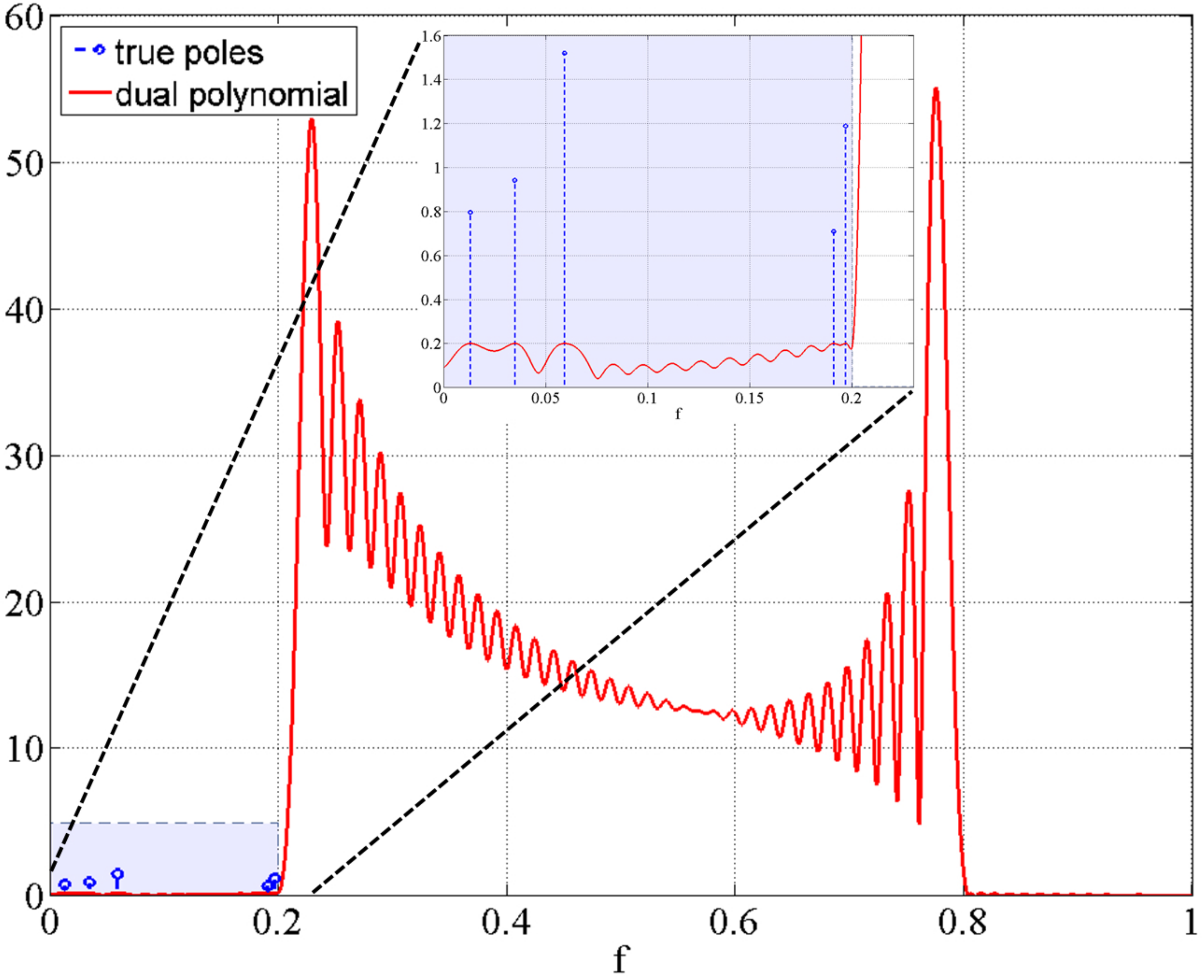}%
  \label{fig:with_probpriors}%
}
\caption{\small Frequency localization using dual polynomial for $\{n, s, m\}$ $=$ $\{64, 5, 64\}$. The probabilistic priors are $p_F(f)|_{\mathcal{B}_1 = [0, 0.2]} = 4.9801$ and $p_F(f)|_{\mathcal{B}_2 = (0.2, 1]} = 0.005$. The insets show the same plot on a smaller scale.}
\label{fig:recovery_illustprob}
\end{figure}
\begin{figure}[!t] \centering
\includegraphics[width=0.45\textwidth]{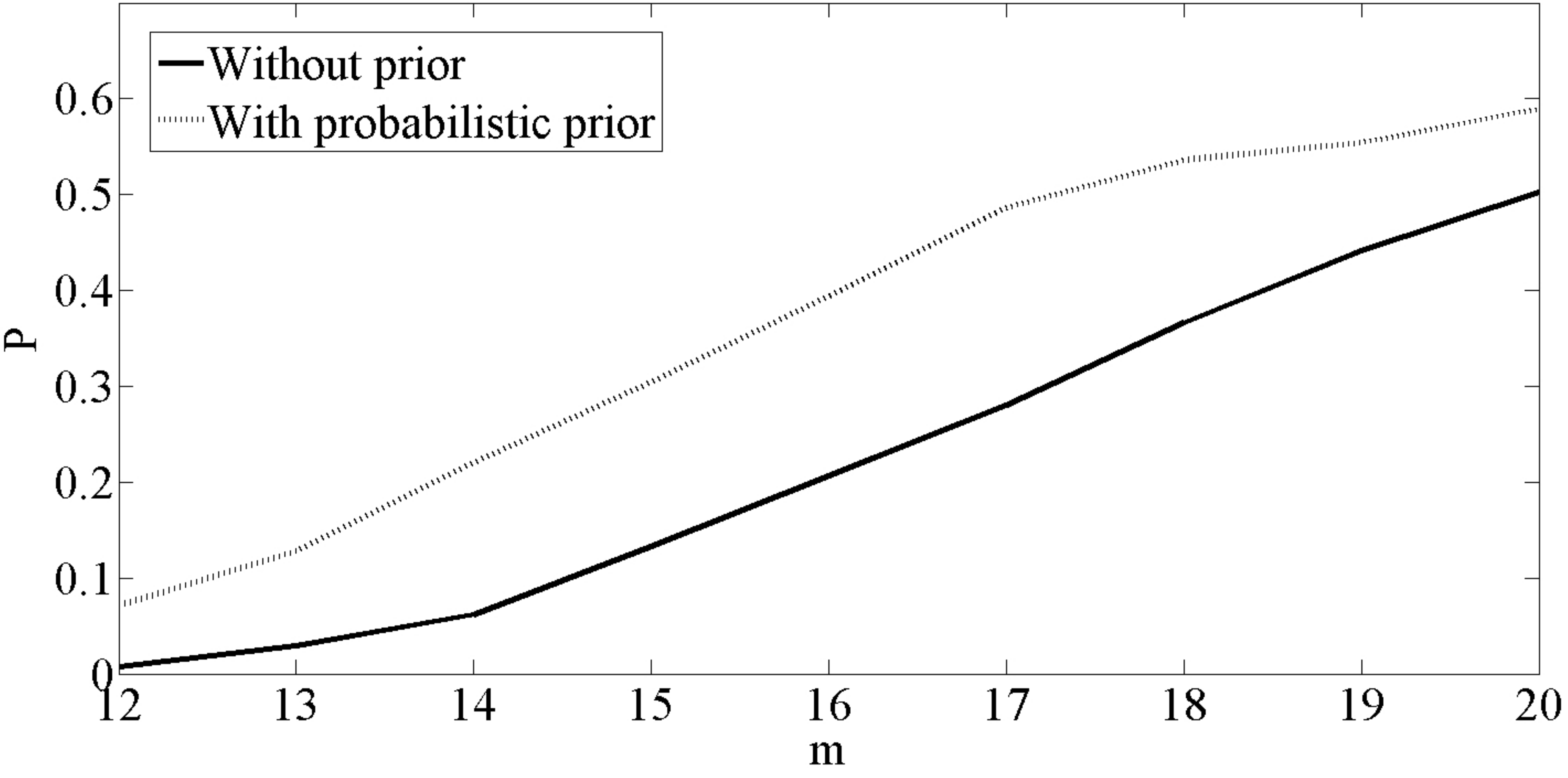}
\caption{\small The probability $P$ of perfect recovery over 1000 trials for $\{n, s\} = \{64, 5\}$. The probabilistic priors are $p_F(f)|_{\mathcal{B}_1 = \{[0, 0.3] \bigcup (0.7, 1]\}} = 0.0025$ and $p_F(f)|_{\mathcal{B}_2 = (0.3, 0.7]} =  2.4963$.}
\label{fig:statcomp_probblock2}
\end{figure}
\begin{figure}[!t]
\centering
\subfloat[Without any priors]{%
  \includegraphics[width=3.95cm]{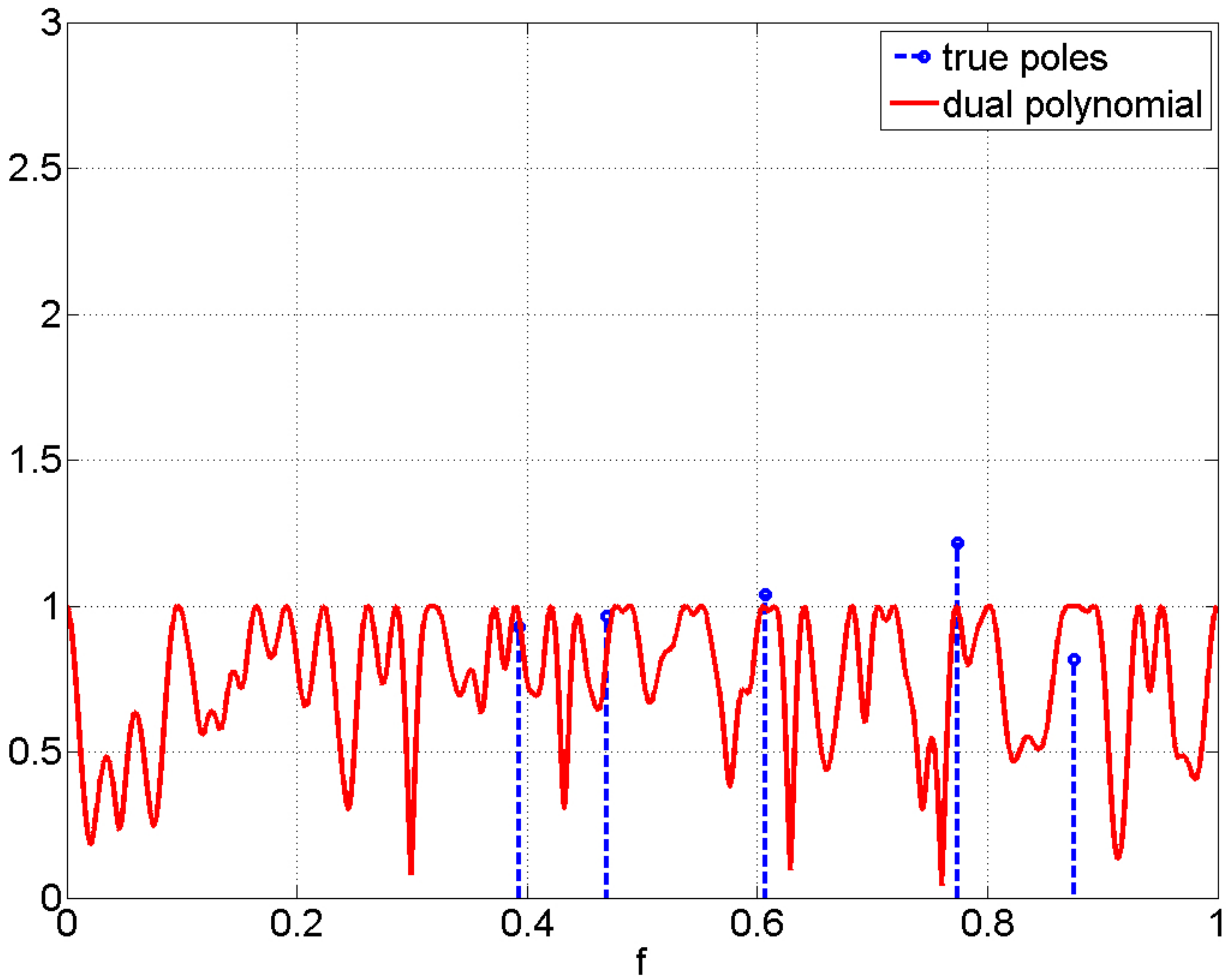}%
  \label{fig:no_blockpriors}%
}\qquad
\subfloat[With block priors]{%
  \includegraphics[width=3.95cm]{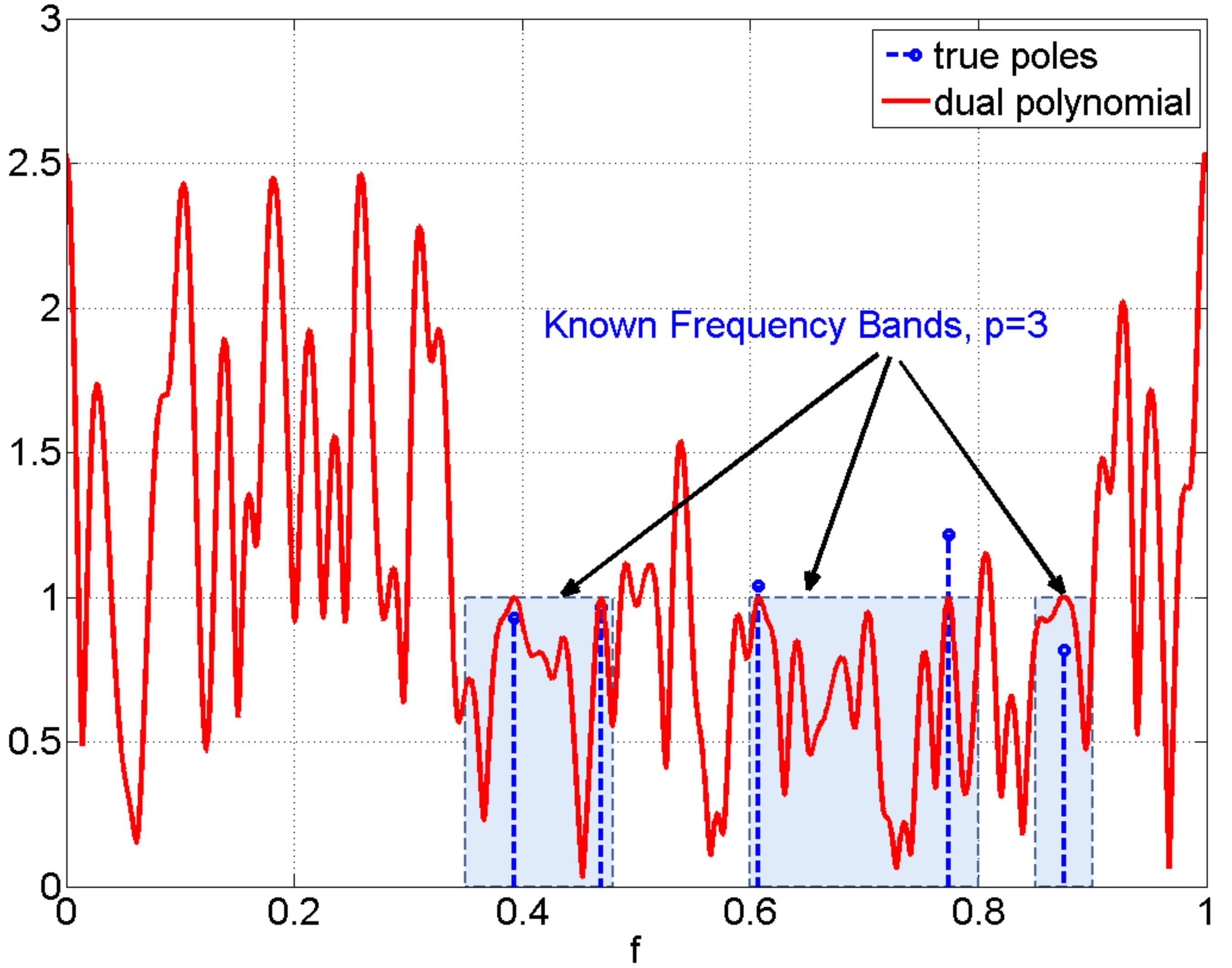}%
  \label{fig:with_blockpriors}%
}
\caption{\small Frequency localization using dual polynomial for $\{n, s, m\} = \{64, 5, 20\}$. The block priors are $\mathcal{B} =$ $[0.3500, 0.4800]$ $\bigcup$ $[0.6000, 0.8000]$ $\bigcup$ $[0.8500, 0.9000]$.}
\label{fig:recovery_illustblock}
\end{figure}
We evaluated our algorithms through numerical experiments using the SDPT3 \cite{tutuncu2003solving} solver for the semidefinite programs. In all experiments, for a particular realization of the signal, the phases of the signal frequencies were sampled uniformly at random in $[0, 2\pi)$. The amplitudes $|c_j|, j = 1, \cdots, s$ were drawn randomly from the distribution $0.5 + \chi^2_1$ where $\chi^2_1$ represents the chi-squared distribution with 1 degree of freedom.
\subsection{Probabilistic Priors}
\label{subsec:numsim_probpriors}
We evaluated the semidefinite program (\ref{eq:probpriorSDP}) for the case when $p = 2$. Here, $\mathcal{B}_1 = [0, 0.2]$ and $\mathcal{B}_2 = (0.2, 1]$ so that $\mathcal{B}_1 \bigcup \mathcal{B}_2 = [0, 1]$. We consider the situation when the probability of occurrence of signal frequency in $\mathcal{B}_1$ is 1000 times higher than $\mathcal{B}_2$. This results in the pdf values of $p_F(f)|_{\mathcal{B}_1} = 4.9801$ and $p_F(f)|_{\mathcal{B}_2} = 0.005$. A suitable sub-optimal choice of $w(f)$ could be simply $w(f) = \nicefrac{1}{p_F(f)}$, so that the associated weights are given by  $w_1 = 0.2008$ and $w_2 = 200.8000$. For each random realization of the signal, the signal frequencies are drawn randomly based on the given probability density function.\\
\textbf{Experiment A.1.} A simple illustration of the signal recovery using (\ref{eq:probpriorSDP}) is shown through frequency localization in Figure \ref{fig:recovery_illustprob}. For a signal of dimension $n = 64$ and number of frequencies $s = 5$, Figure \ref{fig:no_probpriors} shows that even when all samples are observed ($m = 64$), the standard atomic norm minimization (\ref{eq:semiotg}) is unable to recover any of the frequencies, for the maximum modulus of the dual polynomial assumes a value of unity at many other frequencies. However, given the probabilistic priors, semidefinite program (\ref{eq:probpriorSDP}) is able to perfectly recover all the frequencies as shown in Figure \ref{fig:with_probpriors}. Here, $|Q(f_j)| = w_1 = 0.2008$ for $f_j \in \mathcal{F} \subset \mathcal{B}_1$, and $|Q(f_j)| = w_2 = 200.8$ for $f_j \in \mathcal{F} \subset  \mathcal{B}_2$.\\
\textbf{Experiment A.2.} A comparison of the statistical performance of (\ref{eq:probpriorSDP}) with the standard atomic norm for $n=64$ is shown in Figure \ref{fig:statcomp_probblock2} over 1000 trials. Here, the pdf $p_F(f)$ is 1000 times higher in the subband $(0.3, 0.7]$ than the rest of the spectrum. We note that the weighted atomic norm is about twice more successful than the standard atomic norm in recovering the signal frequencies.
\subsection{Block Priors}
\label{subsec:numsim_blockpriors}
\begin{figure}[!t]
\centering
\subfloat[Without any priors]{%
  \includegraphics[width=3.95cm]{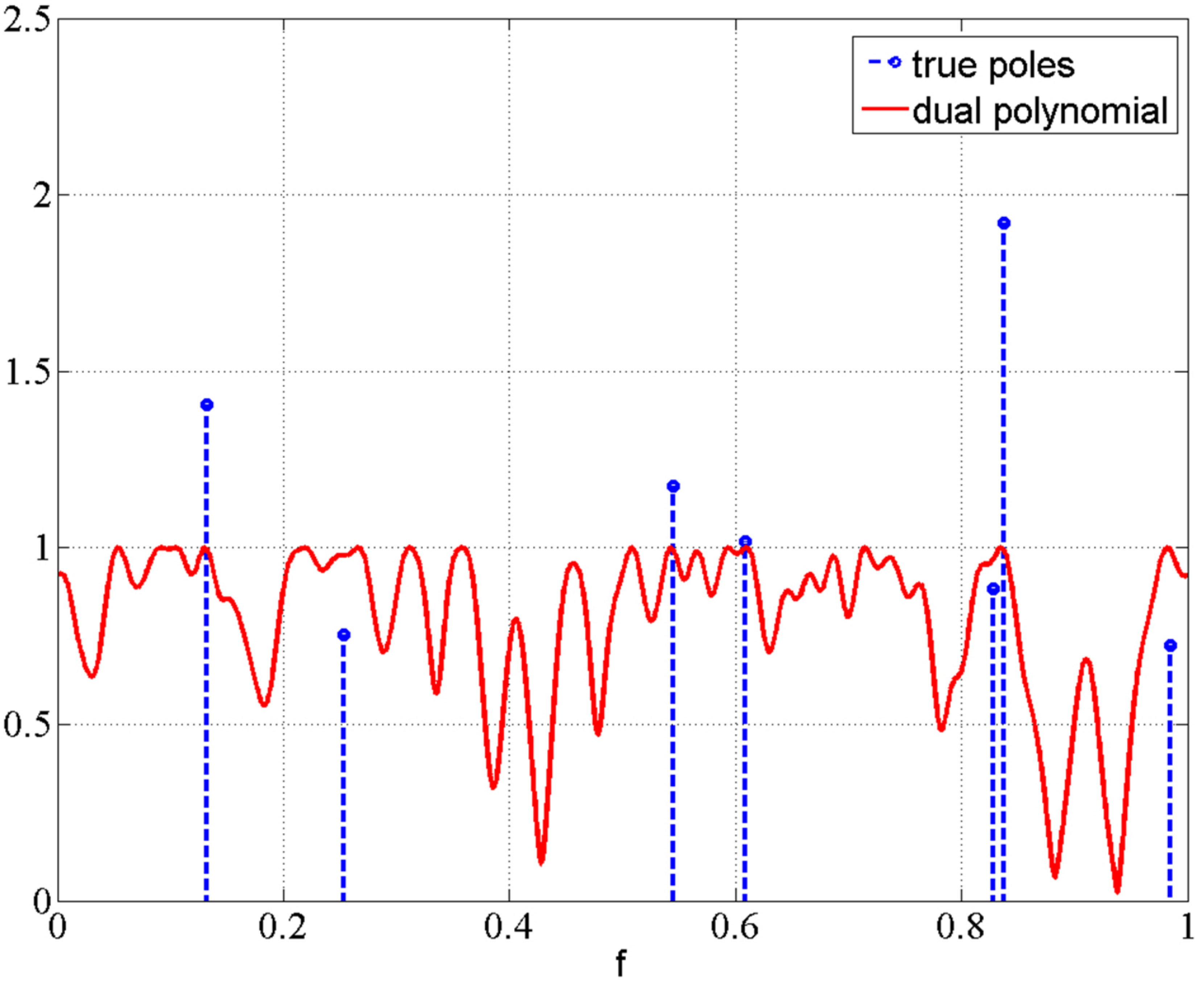}%
  \label{fig:no_smallblockpriors}%
}\qquad
\subfloat[With block priors]{%
  \includegraphics[width=3.95cm]{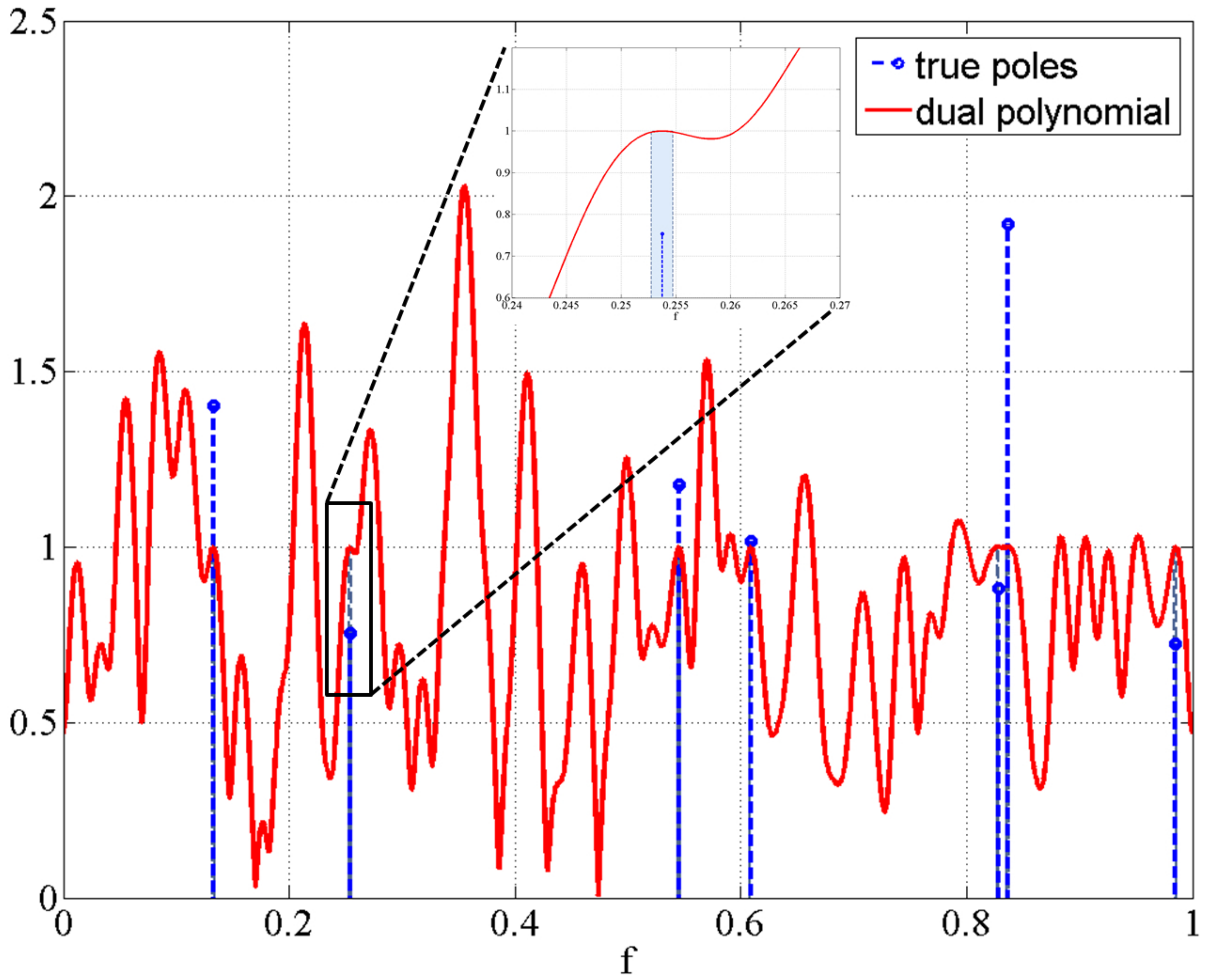}%
  \label{fig:with_smallblockpriors}%
}
\caption{\small Frequency localization using dual polynomial for $\{n, s, m\} = \{64, 7, 18\}$. The block priors consist of small frequency bands around each true pole $f_j$ such that $\mathcal{B} = \bigcup_{k = 1}^{s} \mathcal{B}_k = \bigcup_{k = 1}^{s} [f_j - 0.001, f_j + 0.001]$. The bottom plot has been magnified in the inset to show the size of the block prior.}
\label{fig:recovery_illustsmallblock}
\end{figure}

\begin{figure}[!t]
\centering
\subfloat[Three block priors]{%
  \includegraphics[width=3.5cm]{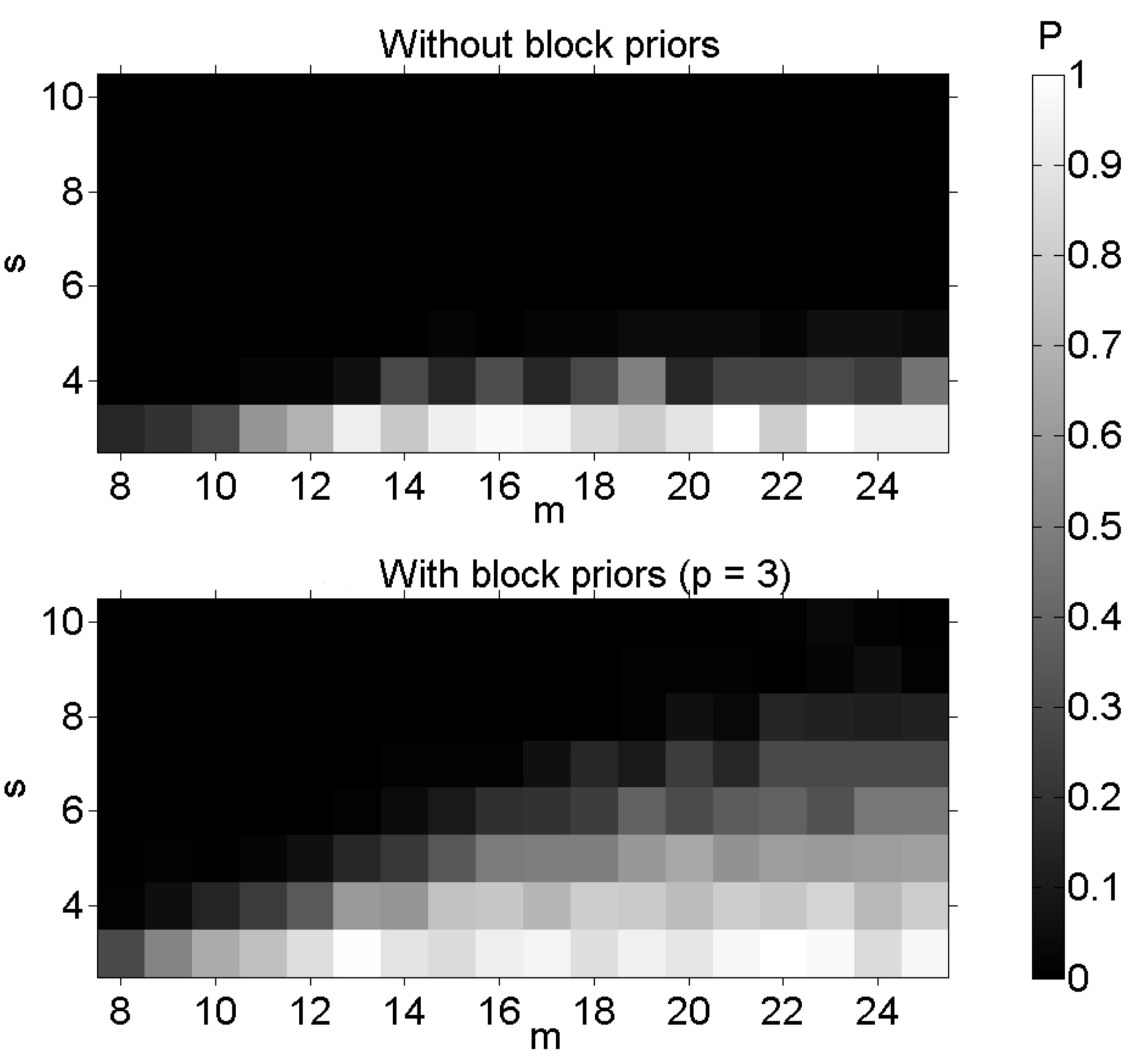}%
  \label{fig:statcomp_block}%
}\qquad
\subfloat[One block prior per pole]{%
  \includegraphics[width=4.6cm]{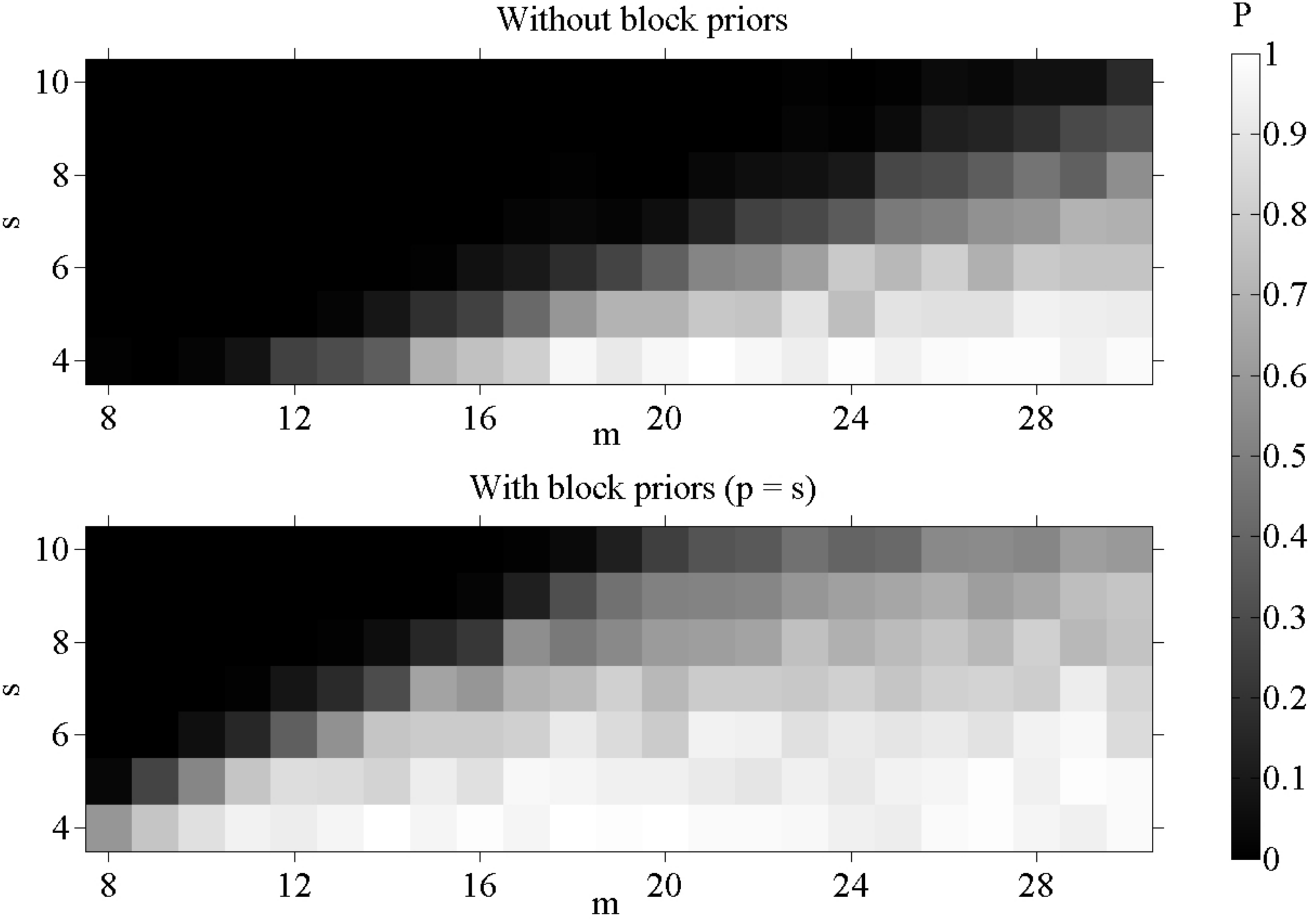}%
  \label{fig:statcomp_smallblock}%
}
\caption{\small The probability $P$ of perfect recovery over 100 trials for $n = 64$. The performance of standard atomic norm is compared with the block prior setups of Figure \ref{fig:recovery_illustblock} (left) and Figure \ref{fig:recovery_illustsmallblock} (right).}
\label{fig:statcomps_blocks}
\end{figure}

We evaluated the performance of spectrum estimation with block priors through numerical simulations for the semidefinite program in (\ref{eq:blocksparsitySDP}). While generating signals in these simulations, the frequencies are drawn uniformly at random in the set of subbands $\mathcal{B} = \bigcup_{k = 1}^{p} \mathcal{B}_k \subset [0, 1]$.\\
\textbf{Experiment B.1.} We first illustrate our approach through an example in  Figure \ref{fig:recovery_illustblock}. Here for $n = 64$, we drew $s = 5$ frequencies uniformly at random within $p = 3$ subbands in the domain $[0, 1]$ without imposing any minimum separation condition. Here, $\mathcal{B} =$ $(0.3500, 0.4800)$ $\bigcup$ $(0.6000, 0.8000)$ $\bigcup$ $(0.8500, 0.9000)$. A total of $m = 20$ observations were randomly chosen from $n$ regular time samples to form the sample set $\mathcal{M}$. In the absence of any prior information, we solve (\ref{eq:dualtoatomicminimization}) and show the result of frequency localization in Figure \ref{fig:no_blockpriors}. Here, it is difficult to pick a unique set of $s = 5$ poles for which the maximum modulus of the dual polynomial is unity (which will actually correspond to recovered frequency poles). On the other hand, when block priors are given, Figure \ref{fig:with_blockpriors} shows that solving (\ref{eq:blocksparsitySDP}) provides perfect recovery of all the frequency components, where the recovered frequencies correspond to unit-modulus points of the dual polynomial.\\
\textbf{Experiment B.2.} We then give a statistical performance evaluation of our new method, compared with atomic norm minimization without any priors (\ref{eq:dualtoatomicminimization}). The experimental setup and block priors are the same as in Figure \ref{fig:recovery_illustblock} and no minimum separation condition was assumed while drawing frequencies uniformly at random in the set $\mathcal{B}$. Figure \ref{fig:statcomp_block} shows the probability $P$ of perfect recovery for the two methods for fixed $n=64$ but varying values of $m$ and $s$. For every value of the pair $\{s, m\}$, we simulate 100 trials to compute $P$. We note that if the frequencies are approximately known, our method greatly enhances the recovery of continuous-valued frequencies.\\
\textbf{Experiment B.3.} To illustrate our theoretical result of Theorem \ref{thm:theorem_3s}, we now consider the block prior problem when each of the frequencies are known to lie in extremely small subintervals. For the triplet $\{n, s, m\} = \{64, 7, 18\}$, Figure \ref{fig:recovery_illustsmallblock} depicts the frequency localization for a random realization of the signal $x$. In the absence of any prior knowledge, the standard atomic norm minimization of (\ref{eq:semiotg}) fails in locating any of the signal frequencies (Figure \ref{fig:no_smallblockpriors}). However, as shown in Figure \ref{fig:with_smallblockpriors}, if the frequencies are approximately known (or, in other words, the frequency subband of the block prior is very small), then the semidefinite program in perfectly recovers the signal requiring not more than $3s$ number of samples ($m = 18 < 21 = 3s$). In Figure \ref{fig:with_smallblockpriors}, the block priors consist of small frequency bands around each true pole $f_j$ such that $\mathcal{B} = \bigcup_{k = 1}^{s} \mathcal{B}_k = \bigcup_{k = 1}^{s} [f_j - 0.001, f_j + 0.001]$.\\
\textbf{Experiment B.4.} For the same signal dimension, size and number of blocks as in the previous experiment, Figure \ref{fig:statcomp_smallblock} shows a comparison of statistical performance of block prior method with the standard atomic norm minimization over 100 trials. For every value of $s$, the parameter $m$ was varied until $m$ was at least $3s$. We note a considerably higher success rate of block prior method. Please note that the perfect recovery is guaranteed only when the block prior is arbitrarily small.\\
\begin{figure}[!t]
\centering
\subfloat{%
  \includegraphics[width=3.0cm]{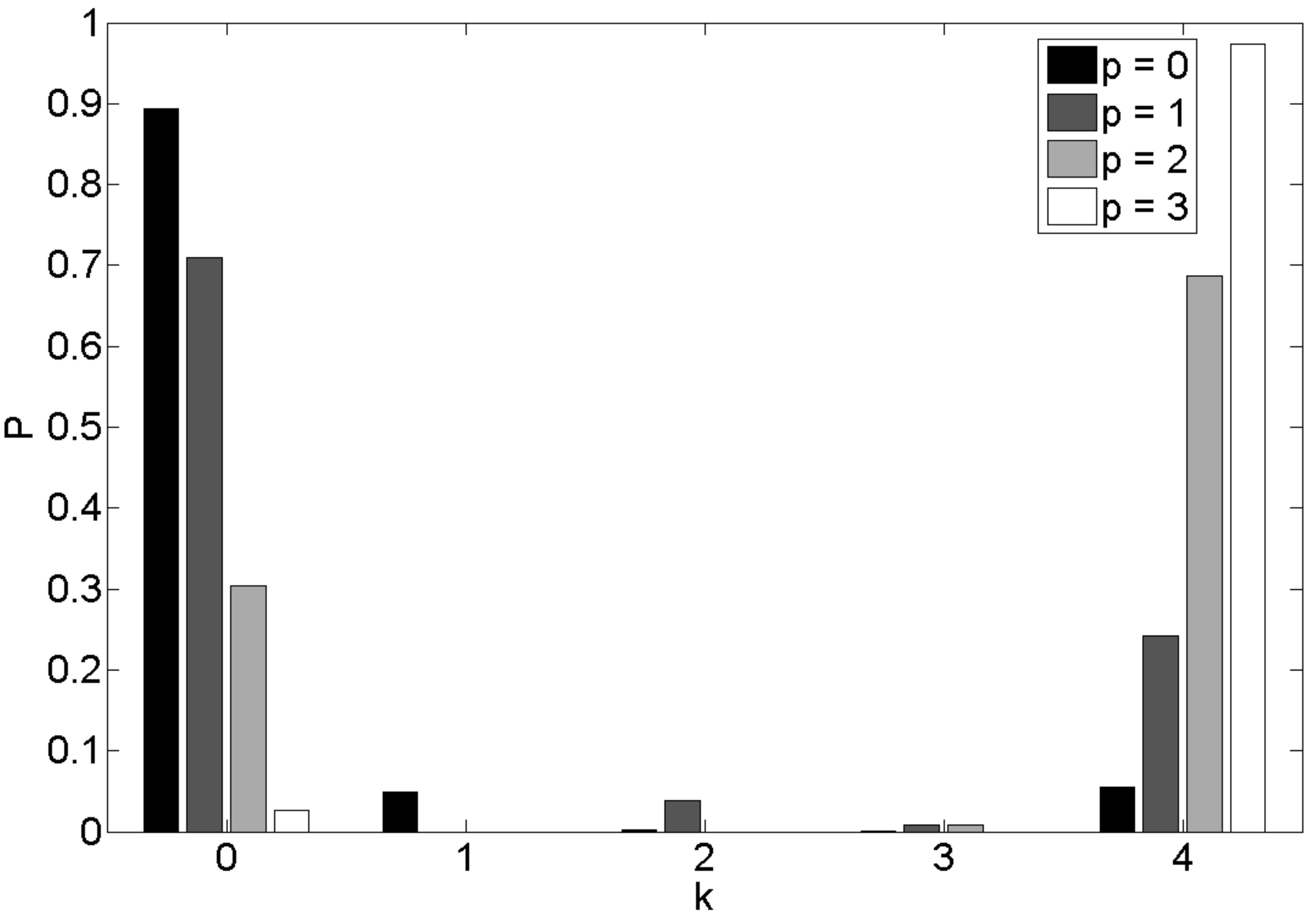}%
  \label{fig:lowdim}%
}\qquad
\subfloat{%
  \includegraphics[width=5.1cm]{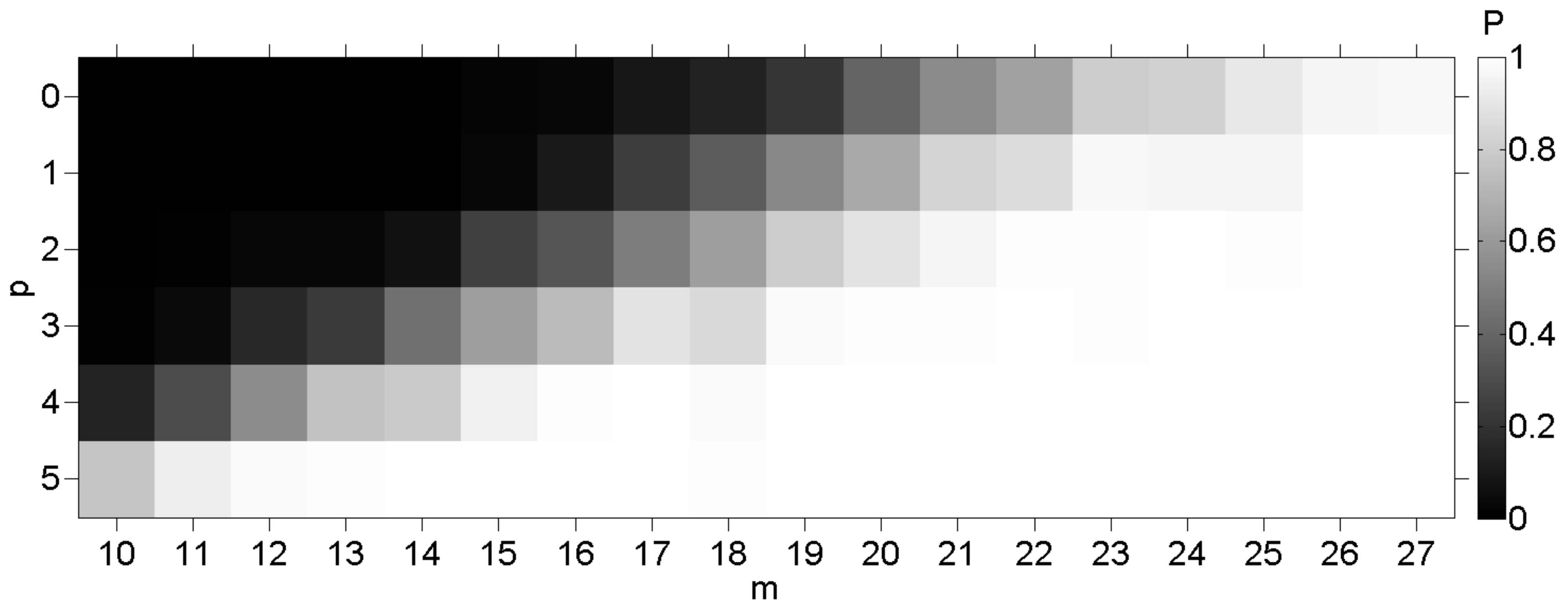}%
  \label{fig:randomset}%
}
\caption{\small The probability $P$ of recovering the unknown spectral content. The probability is computed for 1000 random realizations of the signal for the triple $(n, m, s) = (32, 9, 4)$. (For $k > 0$, $k \le p$ being the invalid cases, the corresponding bars have been omitted.) (b) A higher probability $P$ of recovering all the unknown frequency content can be achieved with a smaller number $m$ of random observations using the \textit{known poles} algorithm. The probability is computed for 100 random realizations with $(n, s) = (80, 6)$.}
\label{fig:statcomp_blocks}
\end{figure}
\subsection{Known Poles}
\label{subsec:numsim_knownpoles}
\begin{figure} \centering
\includegraphics[width=0.45\textwidth]{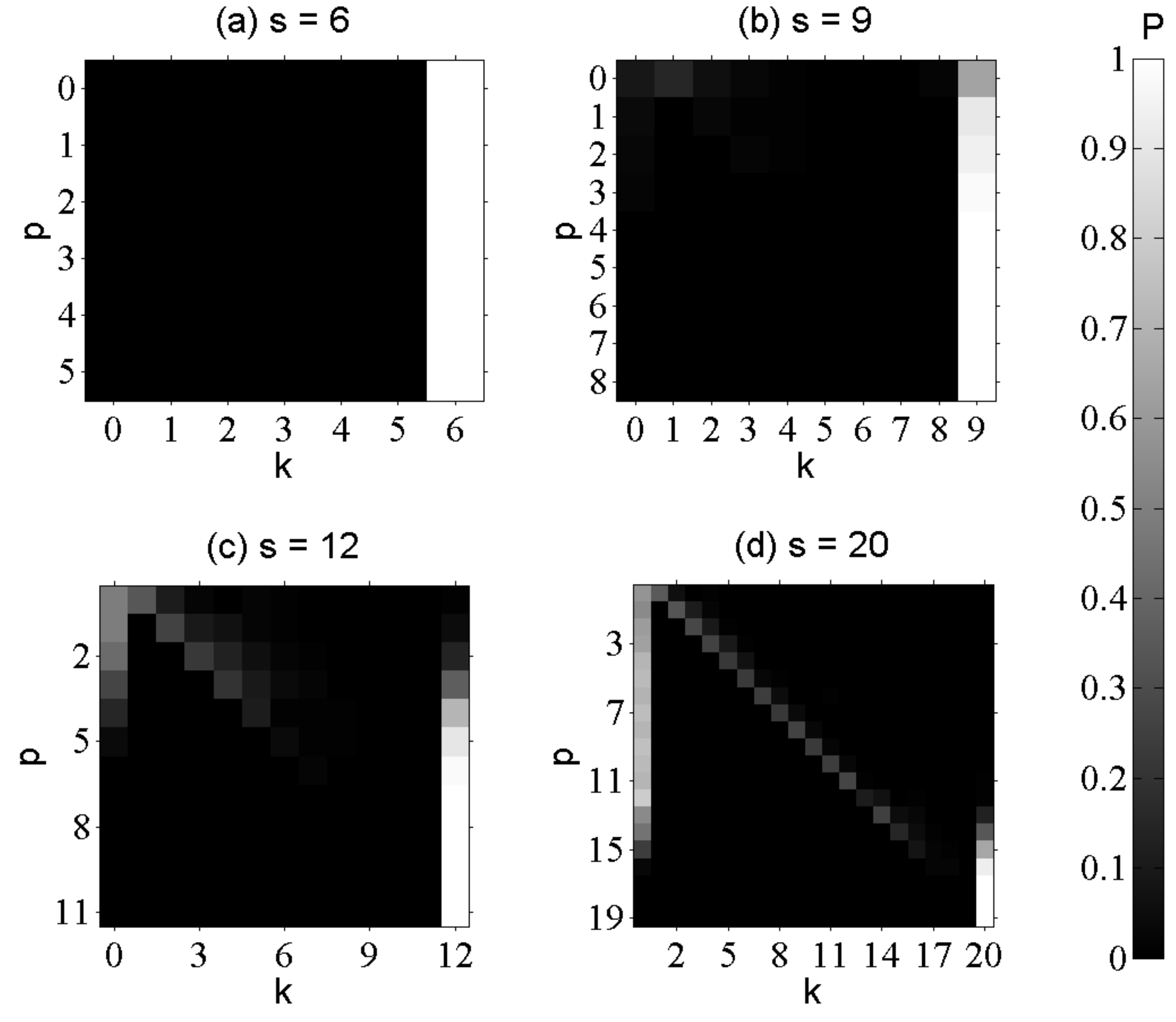}
\caption{\small The probability $P$ of recovering the unknown spectral content for selected values of $s$. The probability is computed for 100 random realizations of the signal with $(n, m) = (256, 40)$. (The lower diagonal cases when $k > 0$, $k \le p$ are invalid, and do not contribute to the result.)}
\label{fig:highdim}
\end{figure}

\begin{figure} \centering
\includegraphics[width=0.45\textwidth]{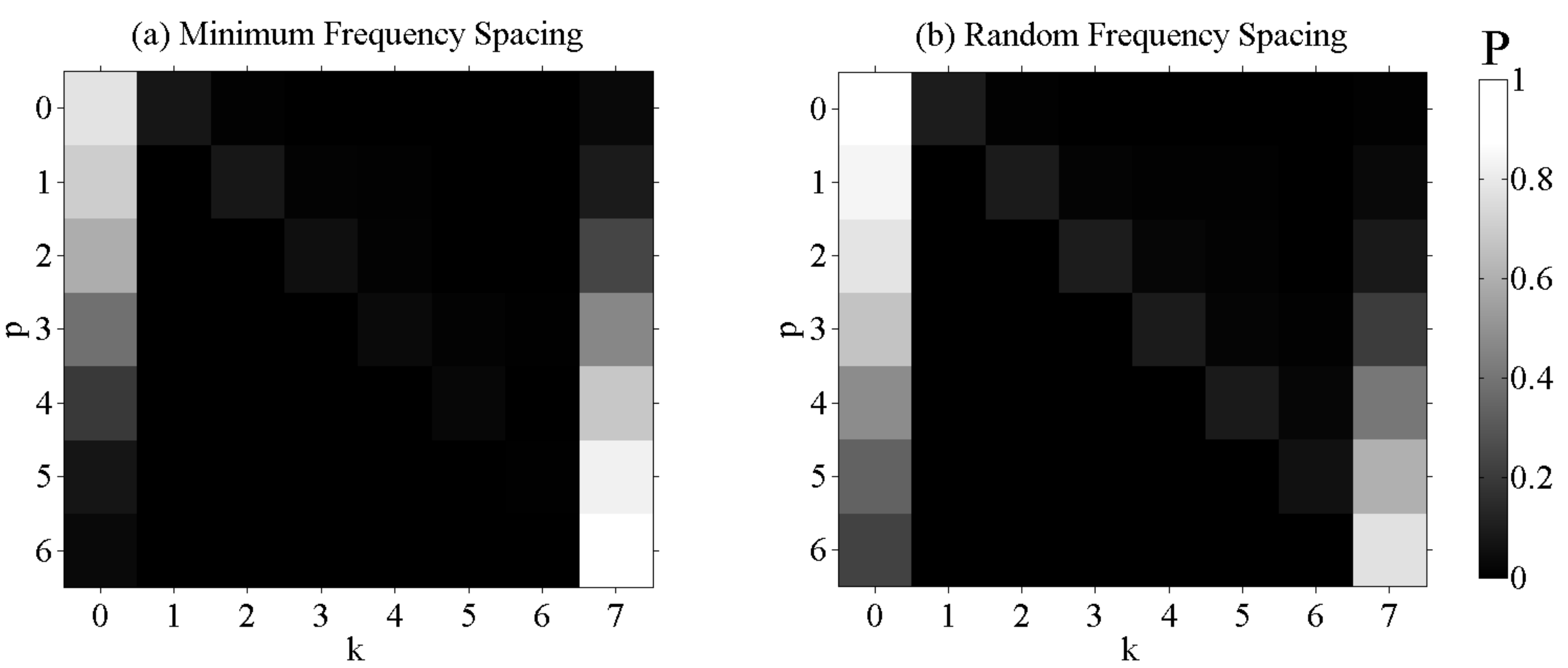}
\caption{\small Performance of the \textit{known poles} algorithm when the frequencies do not satisfy any nominal resolution conditions. The probability $P$ of successfully recovering $k$ frequencies is computed for 1000 realizations of the signal with dimensions $(n, m, s) = (40, 15, 7)$. (a) $\Delta f = \nicefrac{1}{(n-1)}$ (b) Frequencies are selected uniformly at random in the band $[0, 1]$.}
\label{fig:freqspacing}
\end{figure}

We evaluated the \textit{known poles} algorithm through a number of simulations to solve the semidefinite program (\ref{eq:semiprior}).  In all our experiments, the $s$ frequencies of the artificially generated signal were drawn at random in the band $[0, 1]$. Except for Experiment 4, the sampled frequencies were also constrained to have the minimum modulo spacing of $\Delta f = \nicefrac{1}{\lfloor (n-1)/4 \rfloor}$ between the adjacent frequencies. This is the theoretical resolution condition for the results in \cite{tang2012csotg}, although numerical experiments suggested that frequencies could be closer, i.e., $\Delta f$ could be $\nicefrac{1}{(n-1)}$. While working with the \textit{known poles}, we draw the first known frequency uniformly at random from the set of $s$ frequencies. As the number $p$ of \textit{known poles} increases, we retain the previously drawn known frequencies and draw the next known frequency uniformly at random from the remaining set of existing signal frequencies.\\
\textbf{Experiment C.1.} We simulated a low-dimensional model with the triple $(n, m, s) = (32, 9, 4)$ and first solved the semidefinite program (\ref{eq:semiotg}) which does not use any prior information, i.e., $p = 0$. For the same realization of the signal, we then successively increase $p$ up to $s-1$, and solve the optimization (\ref{eq:semiprior}) of the \textit{known poles} algorithm. At every instance of solving an SDP, we record the number $k$ of successfully recovered frequencies along with their complex coefficients. This number also includes the known frequencies if the recovery process returns exact values of their complex coefficients. $k = s$ corresponds to \textit{complete success}, i.e., recovering all of the unknown spectral content. $k = 0$ is \textit{complete failure}, including the case when the complex coefficients of the known frequencies could not be recovered. Figure \ref{fig:lowdim} shows the probability $P$ of recovering $k$ frequencies over $1000$ trials. Although the complex coefficients of the known frequencies were unknown, the \textit{known poles} algorithm increases the probability of accurately recovering all or some of the unknown spectral content.\\
\textbf{Experiment C.2.} We repeat the first experiment for the higher-dimensional pair $(n, m) = (256, 40)$ and vary $s$. The probability $P$ over 100 random realizations of the signal is shown in Figure \ref{fig:highdim} for selected values of $s$. We observe that the probability of successfully recovering all the frequencies using the \textit{known poles} Algorithm \ref{alg:freqLocal} increases with $p$.\\
\textbf{Experiment C.3.} Figure \ref{fig:randomset} shows the probability $P$ of \textit{complete success} as a function of $m$ over 100 trials for the twin $(n, s) = (80, 6)$. We note that the \textit{known poles} algorithm achieves the same recovery probability when compared to (\ref{eq:semiotg}) with a smaller number of random observations.\\
\textbf{Experiment C.4.} We now consider these two cases: (a) when $\Delta f = \nicefrac{1}{(n-1)}$, the resolution limit for the numerical experiments in \cite{tang2012csotg}, and (b) when the frequencies are drawn uniformly at random and do not adhere to any minimum resolution conditions. Figure \ref{fig:freqspacing} shows the probability $P$ of recovering $k$ frequencies over 1000 trials for the triple $(n, m, s) = (40, 15, 7)$. We note that the probability of \textit{complete success} with \textit{known poles} suffers relatively little degradation for the random frequency resolutions. These trials include instances when the minimum resolution condition does not hold, formulation in (\ref{eq:semiotg}) shows \textit{complete failure} but the \textit{known poles} algorithm recovers the unknown spectral content with \textit{complete success}.

\bibliographystyle{IEEEtran}
\bibliography{refs}
\end{document}